\documentclass[%
reprint,
nofootinbib,
 amsmath,amssymb,
 aps, 
 prd
]{revtex4-2}

\usepackage{graphicx}
\usepackage{dcolumn}
\usepackage{bm}

\usepackage{graphicx}
\usepackage{hyperref}
\usepackage{xcolor}
\usepackage[normalem]{ulem}

\newcommand{\epsB}{\epsilon_\mathrm{_B}}
\newcommand{\simmain}{{\bf Reference}}
\newcommand{\simshort}{{\bf Short Downstream}}
\newcommand{\simlong}{\textbf{Long Upstream}}
\newcommand{\simshortl}{\textbf{Noisy}}
\newcommand{\simwide}{\textbf{Wide}}
\newcommand{\simnarrow}{\textbf{Narrow}}
\newcommand{\simhot}{\textbf{Hot}}
\newcommand{\simcold}{\textbf{Cold}}

\begin{document}

\preprint{APS/123-QED}

\title{\textbf{Ultra-long simulations of collisionless relativistic shocks in front-comoving frame: evidence for a steady state and its properties.} 
}%

\author{Mikhail Garasev}
\email{Contact author: garasev@ipfran.ru}
\author{Evgeny Derishev}%

\affiliation{%
Institute of Applied Physics, 46 Ulyanova, 603950 Nizhny Novgorod, Russia 
}%

\date{\today}

\begin{abstract}
We present a series of unprecedently long 2D3V PIC simulations of unmagnetized relativistic $e^{-}e^{+}$-pair shocks performed in a front-comoving frame. By implementing a moving-wall boundary condition in the downstream together with continuous injection at the upstream boundary, we maintain a fixed simulation domain size, opening the way to perform substantially longer simulations. Our longest runs extend beyond $100000\,\omega_p^{-1}$, exceeding the duration of the previously published simulations by a factor of several. Across a diverse set of simulations --- varying upstream/downstream lengths, transverse sizes, and particle-per-cell counts ---
we find strong evidence that the shock, {modelled for a long-time in the front-comoving frame,} approaches an asymptotic, time-independent state. 
In the downstream region, the steady state depends only on the upstream temperature at the injection boundary and {remains the same if we change the downstream and upstream lengths.}
The upstream precursor evolves slower and retains a dependence on the simulation's upstream length, that may be of minor observational consequence, since radiation from astrophysical shocks predominantly originates from the downstream region.
{Our simulations also indicate} that Fermi-type acceleration is limited in energy and a true power-law tail never forms. Another important finding is that the downstream magnetic field has a soliton-like structure, where individual magnetic domains evolve independently, each comprising a compact, highly magnetized core embedded within an extended, weakly magnetized region. The magnetic-field distribution around the centers of these spots has approximately Lorentzian profile.
\end{abstract}

\maketitle

\section{Introduction} \label{sec:intro}

Several types of astrophysical objects --- such as Gamma-Ray Bursts (GRBs), blazars, pulsar wind nebulae, gamma-ray binaries --- form their observed radiation at relativistic collisionless shocks. In most cases, there are only qualitative estimates of parameters in the emitting zone. GRB afterglows are an exception. Thanks to recent breakthrough observations in the TeV spectral domain \citep{Magic_Nature_obs,HESS_Nature_obs,HESS_Science_obs}, it was possible to systematically search the entire parameter space and prove that there is only one region in it, that fits observations \citep{ParameterSpace}. Determination of key shock properties (the strength of self-generated magnetic field, the typical energy of accelerated electrons, and the power-law index of their distribution) with good precision opens the way to check first-principles theories.

At present, there are two competing theoretical models of relativistic shocks. 
The relatively new pair-balance model \citep{PairBalance} describes the shock's front as a result of interplay between radiative (including two-photon pair production) and plasma processes. It predicts the energy of accelerated electrons and positrons in good agreement with observations and provides a qualitative explanation for the long-lived magnetic field. So far no numerical simulation of the pair-balance shock model has been published.
The conventional model claims that the structure of the shock's front is entirely determined by various plasma processes, including kinetic and magneto-hydrodynamical instabilities. It has been extensively studied by means of particle-in-cell (PIC) simulations, whose results we briefly discuss below in this section. 

It was suggested \citep{Sagdeev1966} that collisionless shocks can form even in unmagnetized plasma as a result of Weibel instability \citep{Weibel1959}, which 
may produce rather strong, near equipartition, magnetic fields. 
Later \cite{Medvedev1999, Gruzinov2001} proposed that Weibel instability is the leading mechanism that mediates relativistic shocks, in particular in GRBs. By means of Weibel instability the relativistic shocks convert bulk flow energy into turbulent magnetic field {that further leads to formation of} nonthermal particle distribution {responsible for the observed radiation (synchrotron and inverse Compton)}. The microphysics governing these processes is very complicated and its understanding is based on numerical kinetic simulations. 

The first successful PIC simulations were utilizing the setup with two colliding plasma beams or equivalent setup with the injection of a beam of  plasma into an ambient medium \citep{Silva2003, Nishikawa2003, Frederiksen2004, Hededal2004, Nishikawa2005, Spitkovsky2005, Kato2007}. They have shown the self-generation of the magnetic fields via Weibel instability and rapid formation of the shock front. A simplified setup for relativistic shocks was later developed --- 2-dimensional plasma beam reflecting from a perfectly conducting wall \citet{Spitkovsky2008a}. The results show consistent evidence that unmagnetized relativistic flows can indeed form shocks \citep{Chang2008}. It was found that the magnetization $\epsB$ (defined as the ratio of the magnetic energy density to the total energy density in the hydrodynamical reference frame rises up to $\sim 10\%$–$20\%$ just behind the shock front \citep{Spitkovsky2008a, Chang2008} both for electron-ion and pair plasmas. The self-generated turbulent magnetic field was strong enough to isotropize the particles' velocities. Finally, formation of a clear non-thermal tail in the particle distribution was reported \citep{Spitkovsky2008b}.

Simulations with 2D unmagnetized electron-ion shocks generally confirmed the findings from electron-positron shocks. They have similar structure and amplitudes of the generated fields \cite{Spitkovsky2008a, Bret2016}. 
In an electron-ion plasma, shock formation is a two-stage process. Electrons respond rapidly, on the electron plasma frequency timescale, and form a precursor, but the shock formation is complete only after the ions are decelerated and thermalized. Because the ion frequency is smaller by a factor $\sqrt{m_i/m_e}$, this ion stage is correspondingly slower (here $m_e$ and $m_i$ are the electron and ion masses) \citep{Frederiksen2004, Sironi2011, Sironi2015}. Particle-in-cell simulations of electron-ion flows show that electrons can be substantially preheated by the turbulent fields in the shock precursor, so they enter the downstream region at a temperature not far below that of ions \citep{Sironi2013}.

The results from 3D simulations with electron-ion shocks were not much different from 2D \citep{Nishikawa2003, Nishikawa2005, Hededal2004, Ardaneh2015}. One of the most comprehensive studies to date is \citet{Ardaneh2015},  who performed a 3D simulation of a relativistic electron-ion shock in a setup, where highly relativistic ($\Gamma=15$) flow collides with a plasma at rest. The size of simulation domain was $\sim 100$ ion skin depths, and evolution of the initial state led to formation of forward and reverse shocks mediated by the ion Weibel instability. The results demonstrated that the overall shock structure in 3D is similar to that seen in 2D: there exist filaments, the efficiencies of electron heating and of the magnetic field generation were in line with the results of 2D simulations. The quality of 3D PIC simulations is steadily improving, but they still remain extremely resource-demanding. Most 3D simulations either use artificially low mass ratios $m_i/m_e \ll 1836$ to reduce computational costs, or follow the shock evolution for a very limited time. Nevertheless, the qualitative agreement between 2D and 3D simulations gives confidence that the insights gained from more advanced 2D studies are applicable to realistic 3D shocks. Key phenomena like Weibel-mediated field growth, the early stages of particle acceleration, and magnetic decay all appear in 3D as they do in 2D.

Once it was established that simulations with 2D electron-positron shocks provide a correct picture of the shock's front structure and give a reasonable approximation to the key parameters of emitting zone (the magnetic field strength and the distribution of non-thermal particles), the research focus shifted back to this simple setup, but now the emphasis is on the long-term evolution of the relativistic shocks. 

A series of works \citep{Keshet2009, Sironi2015, Groselj2024} has traced shock evolution to progressively longer time.
The longest simulation to date, reported in \cite{Groselj2024}, that continued for $\sim 26000 \omega_p^{-1}$, where $\omega_p = \left[ 4\pi e^2 n/(m_e\Gamma)\right]^{1/2}$ is non-relativistic plasma frequency for electron-positron plasma in the upstream. (Here $e$ is the elementary charge, $n$ is the total number density of electrons and positrons combined, measured in the downstream-comoving frame, and $\Gamma$ the Lorentz factor of the upstream flow.) This simulation revealed several new trends in shock's evolution: 
the magnetic-field coherence scale at the shock front grows in time, eventually approaching $100$ plasma skin depths ($\lambda_\mathrm{s} = c/\omega_\mathrm{p})$,
the fraction of energy in the quasi-thermal component drops from $\sim 90\%$ at the start of the simulation to $\sim 60\%$ at its end, and this evolution is accompanied by formation of previously unseen suprathermal component, which connects the Maxwellian peak to the high-energy power-law tail. 
They also reported that the downstream magnetic field becomes spot-like. A large fraction of the magnetic energy is contained in small regions with 
magnetization\footnote{The authors actually calculate the normalized magnetic energy density. The relation between two quantities is detailed in Section.\ref{sec:SteadyState}} as high as $\epsB \sim 0.05$, 
whereas the volume-averaged magnetization is much lower, $\langle \epsB \rangle \sim 10^{-3}$.
Even the record-long simulation of \cite{Groselj2024}, in line with shorter simulations, shows no evidence for a steady state and evolution of the shock-front structure continues for as long as the simulation is run. 

With respect to the observational constraints (mostly due to GRB afterglows), the results of PIC simulations of relativistic collisionless shocks can be summarized as follows.
Numerical models have demonstrated formation of a non-thermal tail in the particle distribution downstream of the shock and the anticipated power-law index (assuming a true power-law will eventually form) is $p \approx 2.2 \div 2.5$, that agrees with observations. The energy fraction in accelerated (non-thermal) particles is $10 \div 20 \%$, high enough to explain the observed luminosities. 
The magnetic field strength seen in simulations close to the shock front would also be sufficient to maintain the observed synchrotron radiation.
However, all existing PIC realizations of the conventional model cannot reproduce the long-lived magnetic field and do not actually reach the energy of accelerated particles needed to explain observations. It should be noted that, despite impressive advances in numerical simulations, it was still unclear whether a steady-state shock front solution exists in the conventional model or not. Even the longest available simulations were still evolving in time, and this raised a hope that long enough simulations may produce results closer to what is required by observations. 

Therefore, the question of existence of the steady-state solution for collisionless shocks and determination of its parameters is critical. 
In this paper, we report results from simulations extended to much longer timescales.
This breakthrough was achieved by performing the simulations in the front-comoving frame rather than the standard downstream-comoving frame, which drastically reduces the computational costs of tracking long-term evolution.

The technical details of our setup are given in Sect.~(\ref{sec:Setup}) and Appendices.~\ref{sec:bc},~\ref{sec:MJ_distribution}. 

We find convincing evidence for the existence of a steady state, as discussed in Sect.~(\ref{sec:SteadyState}). The parameters that the shock demonstrates in our steady-state simulations are still very much offset from the values evaluated from observations of GRB afterglows.
This puts into question whether the conventional model of collisionless shocks can be a relevant basis for explaining GRB afterglows.
In addition, we studied shocks with varying upstream temperatures (up to relativistic values), that partially bridges the gap between our simulations and the pair-balance shock, where the upstream's inertia is dominated by relativistic electron-positron pairs. We find that the magnetization across the shock increases with temperature. The correlation length, the maximum particle energy, and the energy content of accelerated particles also increase with temperature.

In Sect.~(\ref{sec:results}) we review the properties of the late-time stationary state for various parameters and compare our results with previous findings of other authors.
Finally, Sect.~\ref{sec:summary} outlines our main findings and addresses possible limitations of different numerical setups.

\section{Simulation setup} \label{sec:Setup}

We performed a series of very long 2D simulations of the relativistic pair shock evolution using the PIC code EPOCH (\cite{Arber2015}). To trace the shock's evolution for as long as possible, we use electron-positron plasma (this reduces the physical timescale) and implement special boundary conditions that allow us to work in the reference frame, which moves at the velocity equal to that of the shock front (the front-comoving frame). They are briefly described in the  Appendix \ref{sec:bc}.

Traditionally, simulations of relativistic shocks are performed in the downstream-comoving frame. In this setup, the lengths of both the upstream and downstream regions grow linearly in time, so that the computational cost is proportional to the square of simulation time.
In the front-comoving frame, used in this work, it is possible to fix the length of the simulation box, therefore the computational cost grows as the first power of time.

\squeezetable
\begin{table*}
\caption {\label{tab:table1} Table with major parameters of different simulations. Fields are: 'Simulation' - the short name of simulation used in text, $L_\mathrm{ds}$ - the spatial size of downstream, $L_\mathrm{up}$ - the spatial size of upstream, $L_y$ - the spatial size in transverse direction, $T_\mathrm{up}$ - initial temperature of upstream flow in the self reference frame, $\beta_\mathrm{wall}$ - the speed of right boundary in the units of the speed of light, $t_\mathrm{sim}$ - total time of simulation, $n_\mathrm{ppc, ds}$ - the average number of particles in one cell in the downstream on the steady-state part of the simulation {and $n_\mathrm{ppc, up}$ - the average number of particles per cell injected at the upstream boundary.}
}
\begin{ruledtabular}
\begin{tabular}{p{4cm}p{1.3cm}p{1.3cm}p{1.3cm}p{1.3cm}p{1cm}p{1.4cm}p{1cm}p{1cm}}
Simulation name & $L_\mathrm{ds}\omega_p/c$ & $L_\mathrm{up}\omega_p/c$ & $L_{y}\omega_p/c$ & $T_\mathrm{up}/m_ec^2$ & $\beta_\mathrm{wall}$ &  $\omega_p t_\mathrm{sim}$ & $n_\mathrm{ppc, ds}$  & {$n_\mathrm{ppc, up}$}  \\    \hline
\simmain & $1600$ & $2300$ & $400$ & 1 & 0.481 & 105000 & 732 & 354\\ 
\simshort & $800$ & $2300$ & $400$ & 1 & 0.481 & 62000 & 428 & 206 \\ 
\simlong & $1600$ & $4600$ & $400$ & 1 & 0.481 & 68000 & 216 & 104 \\ 
\simnarrow & $1600$ & $2300$ & $200$ & 1 & 0.481 & 55000 & 392 & 190 \\ 
\simwide & $1600$ & $2300$ & $1000$ & 1 & 0.481 & 44000 & 92 & 44 \\ 
\simcold & $1600$ & $2300$ & $400$ & 0.1 & 0.498 & 81000 & 384 & 192\\ 
\simhot & $1600$ & $2300$ & $400$ & 5 & 0.468 & 62000 & 376 & 180 \\
\simshortl & $1600$ & $2300$ & $400$ & 1 & 0.481 & 67000 & 48 & 24
\end{tabular}
\end{ruledtabular}
\label{SummaryTable}
\end{table*}

We performed several simulations, whose parameters are summarized in Table~(\ref{SummaryTable}).
Following \cite{Chang2008}, we use inverse upstream plasma frequency $\omega_\mathrm{p}^{-1}, \,\omega_\mathrm{p}^2 = 4\pi e^2n_0/(m_e\Gamma)$  as the timescale and plasma skin depth as the spatial scale. Here $n_0$ is the total (electron + positron) plasma density of unperturbed upstream flow, measured in the front-comoving frame. In all simulations, the cell size is $\Delta x = 0.38\lambda_\mathrm{s}$ and the time step is $\Delta t = 0.19\,\omega_\mathrm{p}^{-1}$. To reduce numerical noise, we use cubic spline shapes for particles and apply a low-pass filter for the electric currents at each time step. 

{We test the adopted resolution by comparing the results for \simmain{} run with a shorter simulation on a finer grid with $\Delta x = 0.19 c/\omega_p$ and find no statistically significant difference.}

Simulation time exceeds $t_\mathrm{sim} = 100000\,\omega_\mathrm{p}^{-1}$ in our best runs, about four times as long as the longest so far simulation of \cite{Groselj2024}. We focus primarily on the \simmain{} simulation. For that particular run, initially, 60\% of the box is filled with the upstream plasma and 40\% with the shocked downstream plasma. The simulation continues up to $t_\mathrm{sim} = 105000\omega_\mathrm{p}^{-1}$ with about $5\times10^{9}$ particles in each species. 

In our simulations the shock moves along the $x$-axis. 
Initially, the simulation box is divided into two parts, the upstream and the downstream part. The position of the shock front is defined as a point where the density increases halfway between the upstream and downstream values.
Both parts are populated with electron-positron plasma without any electromagnetic fields. 
In the upstream, the momentum distribution for both particle species is a {3D} Maxwell-J\"{u}ttner distribution, Lorentz-boosted to have hydrodynamic Lorentz factor $\Gamma = 24.6$ (this corresponds to $\Gamma_\mathrm{ds} \approx 15$ in the downstream frame). 
The algorithm {that has been used in current work} for generating such a distribution is given in Appendix.~\ref{sec:MJ_distribution}. {There are alternative algorithms for loading Lorentz-boosted Maxwell-J\"{u}ttner distributions, e.g. \citet{Zenitani2015}. }
We used several values for the comoving-frame temperature in the upstream (see Table~\ref{SummaryTable}).
The high, relativistic upstream temperature serves a dual purpose: it inhibits the generation of unphysical electromagnetic modes {(which mainly arise due to development of numerical Cherenkov instability (NCI))} far upstream and approximates the pair-balance shock, where relativistic electron-positron pairs dominate the upstream inertia.

In the downstream, the initial distribution function is also boosted Maxwell-J\"{u}ttner. The boost velocity is adjusted to match the downstream velocity (see Table~(\ref{SummaryTable}) for the values). 
The initial downstream temperature and density are chosen to ensure continuity of energy and particle number fluxes at the boundary between the upstream and downstream parts. 
Note that the actual shock velocity does not exactly satisfy the jump conditions for a plane-parallel shock. The small difference is due to energy and momentum carried by escaping particles into the upstream and by the magnetic field into the downstream.

For electromagnetic fields we employ open boundary conditions both in the upstream and in the downstream. We also use open boundary condition for outgoing particles in the upstream --- they are removed once they leave the simulation box. Concurrently, at every time step we inject new particles at the upstream boundary as if the flow with initial parameters extended further into the upstream. The new particles replace those that moved towards the shock front and are injected in pairs --- for every electron we inject a positron with the same momentum and at the same location. This is done to reduce noise generation at the upstream boundary.

For particles at the downstream boundary we use the moving wall boundary condition, which is described in App.~(\ref{sec:bc}).  
In effect, this boundary condition is equivalent to the reflection from a wall that moves at some velocity with respect to the simulation frame. We choose this velocity equal to the bulk velocity of downstream. 
In the transverse direction  ($y$-axis) we use periodic boundary conditions for both particles and fields.

Since evolution of the initial state into the shock-like structure takes several (3-4) light-crossing times being of little physical interest, we start all our simulations with a reduced number of particles (having the correspondingly greater weight) and then increase the particle load step by step. We make a fork in one of the simulations, continuing it both with initial and increased number of particles per cell. In this way, we check how the shock structure depends on numerical noise.

{To test the consistency of the setup we have performed two types of control tests. First, we propagated the injected distribution with the electromagnetic field-particle interaction disabled. With the inflow and wall velocities matched, the bulk flow remains stationary and we do not observe any changes in the flow near the boundaries. This test isolates the boundary implementation itself.}

{Second, with the self-consistent electromagnetic interaction restored, a gradual deceleration of the relativistically drifting upstream appears. This is infamous obstacle appeared in simulations of relativistic flows, due to the development of NCI. Its magnitude depends strongly on the injected Lorentz factor, upstream temperature, and number of particles per cell, as expected for a numerical instability associated with the relativistic drift. For the numerical parameters of the \simmain{} run and for \(n_{\rm ppc,up}=128\), the bulk Lorentz factor gradually decreases approximately from \(\Gamma\simeq25\) to \(\Gamma\simeq 22\) over the length of the upstream during the time required to reach the front. }

\section{{Long-term evolution and formation of the steady state}} \label{sec:SteadyState}

For every simulation from the list in Table~\ref{SummaryTable}, we continuously monitor several characteristics of the shock, both globally and as functions of coordinate along the flow. 
To illustrate our analysis in this section we choose four of the characteristics (see below), although a different set results in a similar overall picture and the same conclusions. The particular choice was largely guided by convenience of comparison with other works in the field.

\begin{figure}
   \centering
   \includegraphics[width=1.0\columnwidth]{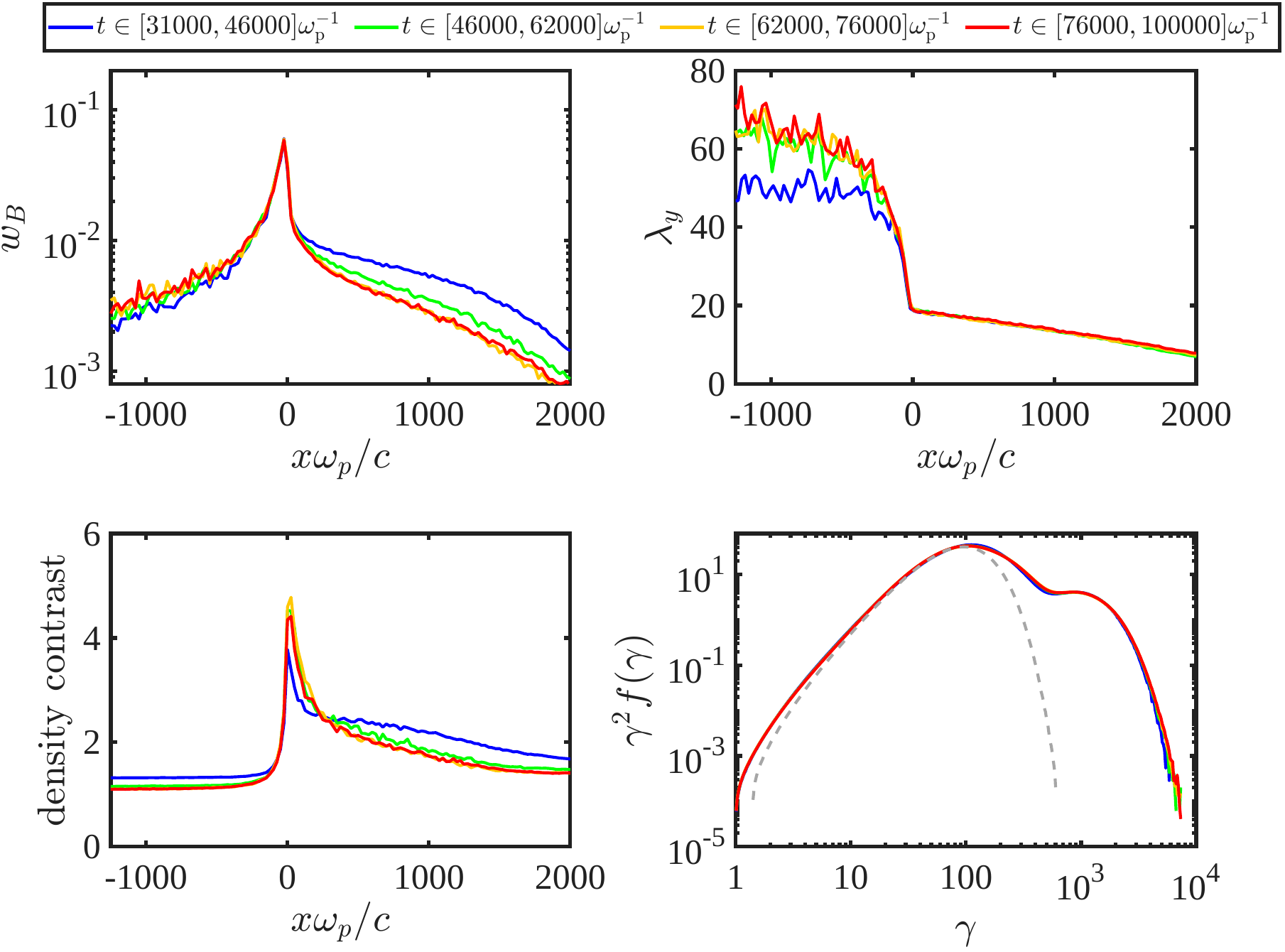}  
   \caption{Evolution of normalized magnetic energy density, correlation length, density contrast as a function of coordinate along the flow and particle energy distribution in the downstream region ($-750<x\omega_\mathrm{p}/c<-500$) in the \simmain{} simulation. Each curve is time averaged, the averaging interval showed in legend. As one can see, magnetic energy, correlation length and density contrast demonstrate slowing evolution up to $\omega_\mathrm{p}t\sim 60000$ and are almost constant after that. The particle energy distribution reaches steady state very early in the simulation. Dashed gray line represents the thermal fit to the low energy part of the particle distribution.
   }
   \label{fig_s1}
\end{figure}

\begin{figure}
   \centering
   \includegraphics[width=1.0\columnwidth]{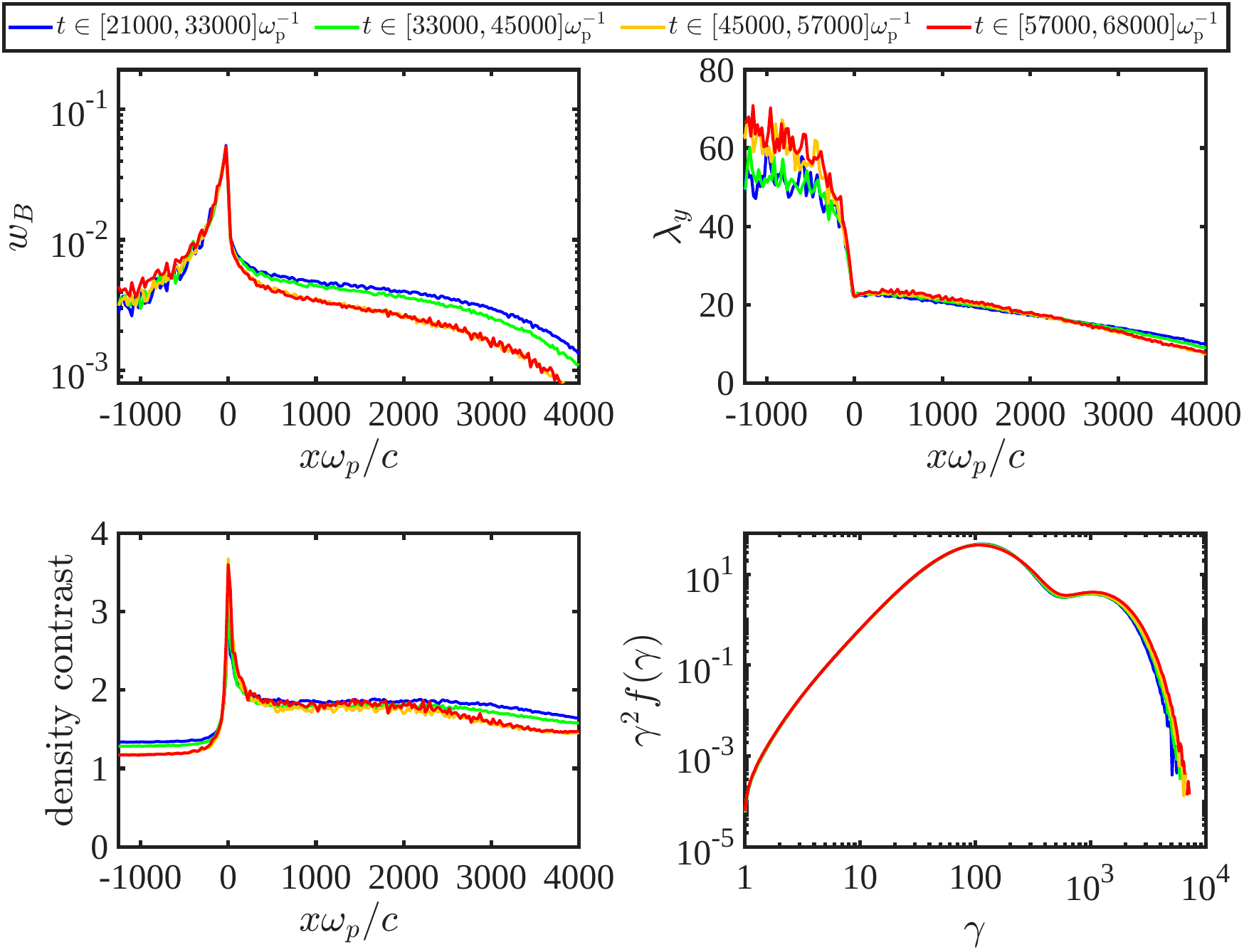} 
   \caption{The same as Fig.\ref{fig_s1} but for \simlong{} run.
   }
   \label{fig_s2}
\end{figure}

Figures \ref{fig_s1} and \ref{fig_s2} show temporal evolution of the shock for two simulations with different upstream length. Each figure contains four panels, with data averaged over time intervals indicated on the plots. The first panel in each figure shows the spatial profile of $y$-averaged mean of the normalized magnetic energy density, 
\[
w_\mathrm{_B} = \left\langle \frac{B^2}{8\pi n_{0}E_0} \right\rangle,
\]
where the average is taken over a region spanning the full $y$-width and extending $25\lambda_s$ along $x$-axis.
Here both the magnetic field and the mean energy of upstream particles at the moment of injection, $E_0$, are measured in the simulation frame. 
In the downstream this quantity closely resembles the magnetization, keeping in mind the density jump and reference frame conversions $w_\mathrm{_B}$ is about $2$ times greater than $\epsB$ in that region. The second plot displays the evolution of the transverse correlation length, defined (following \cite{Groselj2024}) as
 $\lambda_y = \pi\int k_y^{-1}F_{B}dk_y/\int F_{B}dk_y$, 
where $k_y$ is the transverse wavenumber and $F_B = |\hat{B_z}(k_y)|^2$ is the one-dimensional magnetic energy spectrum as a function of $k_y$. 
The third panel shows the {energy distribution over particles' spectrum, $\gamma^2 f(\gamma)$, where normalization of the particle distribution function is $\int f(\gamma) \, d\gamma = 1$.}

The fourth panel depicts the density contrast. 
This quantity characterizes local inhomogeneity of the flow and is evaluated in the following way.
First, during the averaging period, we collect all electron and positron density values in each simulation cell and assign each cell a coordinate equal to its position relative to the shock front (along x-axis). 
Second, we select all cells from the interval of width $25\lambda_\mathrm{s}$ centered at a given coordinate.
Third, we sum up the number of particles in 10\% of cells with the highest densities ($N_\mathrm{high}$) and the number of particles in 10\% of cells with the lowest densities ($N_\mathrm{low}$).  The density contrast is the ratio $N_\mathrm{high}/N_\mathrm{low}$.

In the {\simmain\,  simulation} all the parameters {and their distributions} remain almost constant after $t \gtrsim 50000\,\omega_p^{-1}$, while for {\simlong\, simulation (with two times longer upstream)} we still see the evolution in the magnetic energy density, density contrast and correlation length profiles up to the end of simulation, which lasts for $68000\,\omega_\mathrm{p}^{-1}$. However, this prolonged evolution continues only in the upstream, while changes in the downstream towards the end of simulation are subtle. Note that the particle spectrum in both simulations reaches its steady state relatively fast.
To quantify the rate of transition to the steady state, we compute the time-dependent deviation
\[
\Delta F_\gamma(t) = \int_\gamma (\ln{f_\gamma(t)}-\ln{f_{\gamma, \mathrm{s}}})^2d\gamma, 
\]
where $f_{\gamma}(t)$ is the particle distribution function at time $t$ and $f_{\gamma, 
\mathrm{s}}$ is the final steady-state spectrum (end of simulation). Only Lorentz factors for which $f(\gamma)>f_\mathrm{thr}$ are included; the threshold is chosen so that, on average, each bin contains more than 3 computational particles reducing shot-noise contamination from the distribution's tail. The values $\Delta F(t_k)$ at discrete times are then fitted with \[\Delta F_k =  C_1 + C_2\cdot\exp\left(-t_k/t_\mathrm{s}\right),\] where $C_1$ and $C_2$ are constants and $t_\mathrm{s}$ is the characteristic relaxation time. For all simulations, the fitted $t_\mathrm{s}$ is of the order of a few light-crossing times of the simulation box.

Not only in the two simulations described in detail above, but in all our simulations and for every characteristic we observe qualitatively similar behavior: prominent evolution at start of a simulation gradually slows down and, in most cases, reaches a state where further evolution is indiscernible. We interpret this apparently unchanging final condition as an indication of a steady state. Various characteristics reach the steady state at a varying time. The first to stop evolving is particle distribution in the downstream, and the last is the spatial profile of the normalized magnetic-energy density in the upstream. In fact, the latter distribution continues to evolve till the very end of runs with long upstream and reaches steady state only in the longest runs with short upstream.

\begin{figure}
   \centering
   \includegraphics[width=1.0\columnwidth]{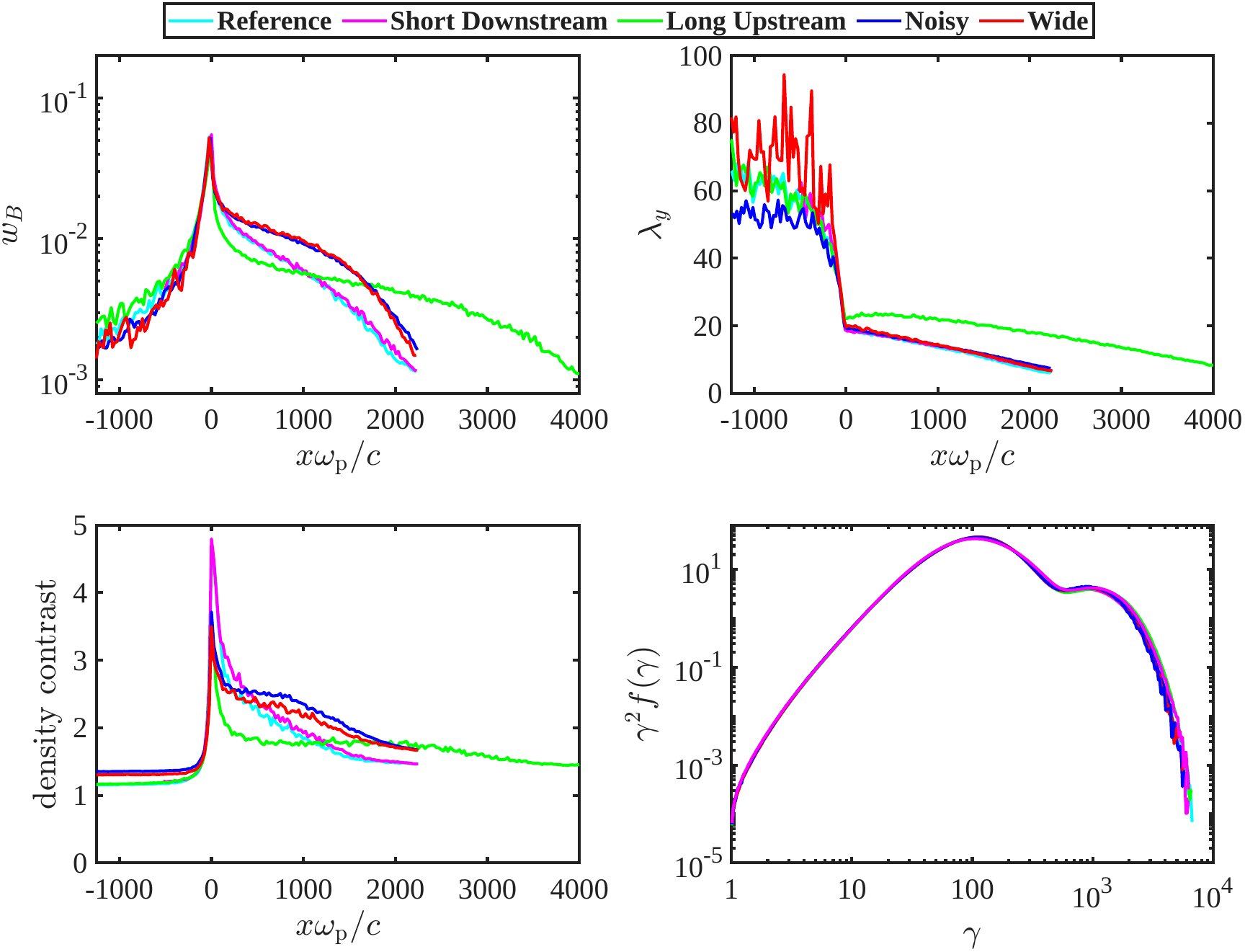} 
   \caption{Late-time profiles of the magnetic energy density $w_\mathrm{_B}$, correlation length $\lambda_y$ and density contrast as a function of coordinate along the flow and particle energy distribution for 5 simulations --- \simmain, \simshort, \simshortl, \simlong, and \simwide{} (see Table~\ref{SummaryTable} for summary of their parameters). 
   }
   \label{fig_s3}
\end{figure}

Our next step is to check whether the late-time stationary state that we observe characterizes physical solution or a particular numerical setup. 
In Fig.~\ref{fig_s3} we compare this state (using the same set of four plots) for several simulations, which have the same physical upstream parameters (temperature, density, hydrodynamic velocity) but different numerical realization. We include \simmain, \simwide, \simlong, \simshort, and \simshortl\, simulations from Table~\ref{SummaryTable}. 
In the downstream, both the particle distribution and the spatial structure of all the parameters that we monitored (not only those shown in Fig.~\ref{fig_s3} demonstrate a close match between the simulations.
On the other hand, we observe differences in the upstream part of the flow, most notably in the magnetic energy density and density contrast as functions of coordinate. 
Changes of the upstream and downstream length, and even a mere increase of numerical noise (in the simulation with fewer particles per cell) --- all result in noticeable differences in the upstream part of the flow.
Despite this fact, we do not see any discernible influence of changes in the upstream on the particle energy spectrum or on the size, shape, and amplitude of the magnetic structures in the downstream.
{Reducing the downstream size (from \(L_\mathrm{ds} = 1600c/\omega_\mathrm{p}\)  to \(800 c/\omega_\mathrm{p}\)) changes the relaxation time, but the measured downstream coherence scale, near-front magnetization,
density contrast and accelerated-particle energy remain almost unchanged. }

To summarize, in all our very long runs we observe the emergence of the steady state for the particle energy spectrum and the magnetic turbulence at the shock front and in the downstream. The time it takes to reach this steady state depends on the simulation frame's spatial extent; typically it takes a few light crossing times across the simulation box. 
Apparent absence of true steady state for the upstream may be due to shortcomings of our approach or may arise from fundamental physical reasons. Interestingly, the existence of a steady state for plane-parallel shock has been recently put into question by \citet{Gruzinov2025}.
In any case, it is the downstream part of the flow that is important for practical applications, such as calculations of broad-band electromagnetic spectra forming at relativistic shocks. Thus, our results provide evidence of the steady-state solutions for a relativistic shock in the practical sense.

\section{Properties of the late-time steady state} \label{sec:results}

Overall, our results (obtained using the front-comoving simulation setup) are in fundamental agreement with the results of previously published simulations of unmagnetized collisionless shocks in the downstream-comoving setup. 
Whenever we observe a qualitative difference in the global flow characteristics, those are not of primary physical importance.  However, we do observe quantitative differences in some key characteristics and qualitative changes in their temporal evolution when the simulations become significantly more evolved compared to previous works. 
We also report several new findings that may contribute to understanding of shock's microphysics.
{When comparing to other works, we mainly} focus on {the paper by \citet{Groselj2024},} which presents the longest {so far} simulation {in the downstream-comoving} setup.

Across all simulations (performed in the front-comoving setup) we observe the same qualitative behaviour: an initial transient phase lasting a few light-crossing times is followed by progressively slower evolution, until the downstream reaches an asymptotic steady-state for all important physical quantities: correlation length, spatial profile of the magnetic energy density, particle energy spectrum, etc. 
The downstream particle spectrum stabilizes earliest, while the slowest convergence is typically associated with upstream precursor quantities (spatial distribution of magnetic energy density, density contrast, and coherence scale). In simulations with an extended upstream region, slow evolution in the far upstream persists for as long as the simulation is continued.

Notably, the downstream steady state appears to be a robust solution. Significant variations in upstream and downstream lengths, transverse size, and particle-per-cell count may alter the far-upstream structure, yet leave the downstream particle spectrum, the magnec energy density distribution and the magnetic-field morphology largely unaffected. This is especially important for practical applications, because the downstream region is the primary emission zone in many relativistic-shock models. 

Formation of such a steady-state has never been observed in the downstream-comoving setup. An apparent reason for this qualitative difference is that the duration of our simulations is at least $2$ times longer than the durations of previously published simulations (the \simmain{} simulation is 4 times longer). Furthermore, reaching a steady state requires 
fully established feedback between the shock front and the reflecting boundary in the downstream, i.e., particles need enough time to undergo several round trips.
In the front-comoving setup, the downstream region does not expand with time, and hence the steady state should settle faster. Indeed, the simulation with reduced downstream length shows a shorter settling time (see Table~\ref{SummaryTable2}). {The detailed discussion of the settling time $t_\mathrm{s}$ is given in the end of section \ref{sec:results}}.

\begin{figure*}
   \centering
   \includegraphics[width=\textwidth]{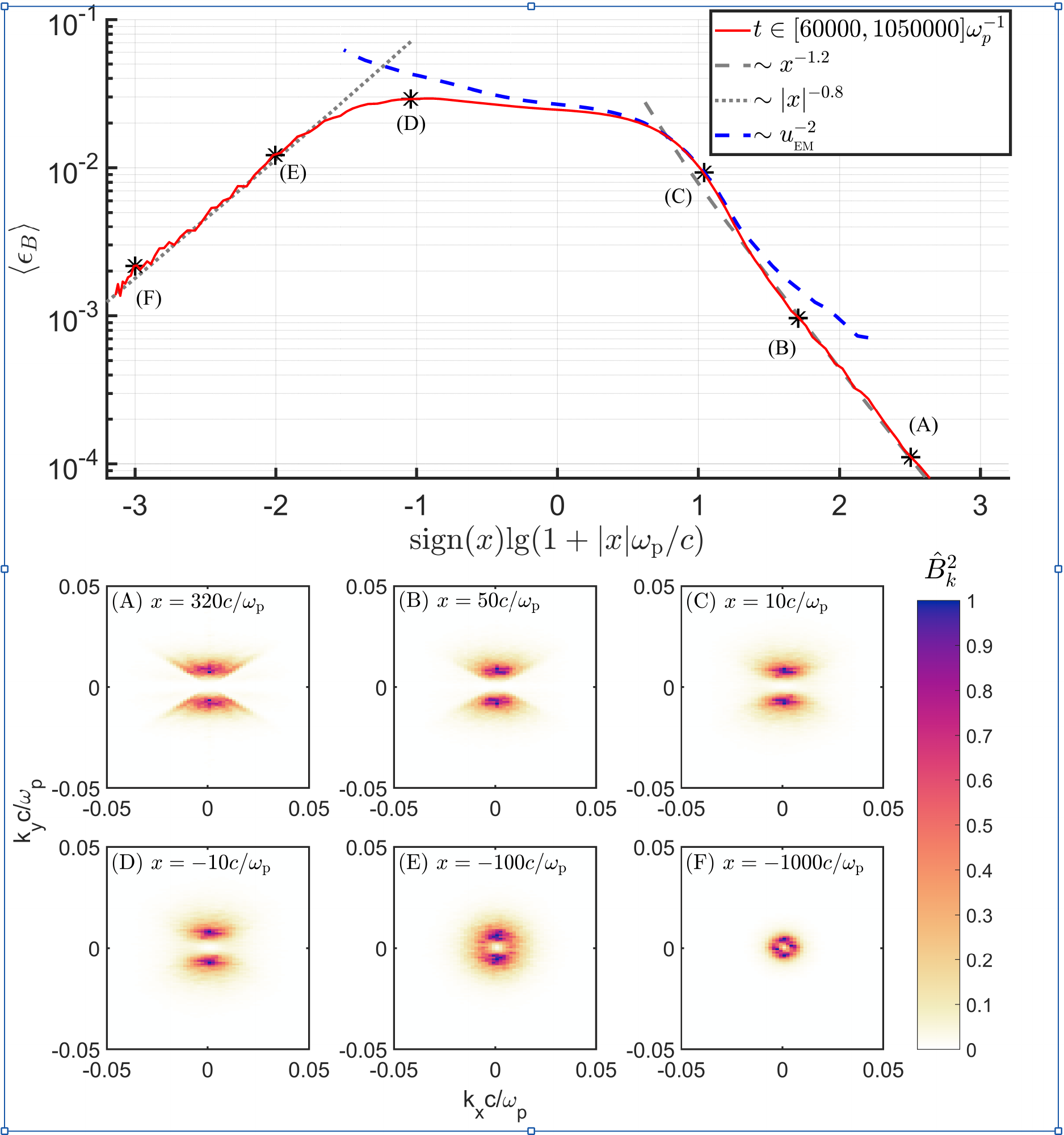}
   \caption{
      {\it Top panel.} Red line -- trasversely and time-averaged magnetization profile $\langle\epsilon_B\rangle$ as function of distance to the shock front, time-averaged over the steady-state interval $t\in(60000, 105000)\,\omega_p^{-1}$ for the \simmain{} simulation. 
      The gray lines are the power-law asymptotics for the upstream ($x>0$, dashed) and the downstream ($x<0$, dotted). 
      Blue dashed line --- the amplification of the magnetic field due to compression of the flow (see the text).
      The black asterisks show the points at which the local spatial power spectra were calculated for panels (A)-(F). 
      {\it Panels (A)-(F):} time-averaged (over $t\in(60000, 105000)\,\omega_p^{-1}$ interval) and normalized spatial power spectra $\hat B_k^2$ at different locations along the flow, shown in the top panel. 
      Each spectrum is calculated in a window of width $\Delta x = 200c/\omega_p$; the windows are centered at $x\in\{-1000, -100, -10, 10, 50, 320\}\,c/\omega_p$ relative to the shock front position. 
      The spectra were calculated in the front-comoving frame (i.e, in the mesh frame) and normalized to a harmonic with the largest amplitude. Further upstream of point (A) the numerical noise becomes noticeable.
   } 
   \label{fig_pp3}
\end{figure*}

Time-averaging over the steady-state part of our simulation allows us to measure magnetization along the flow (see panel (a) of Fig.~\ref{fig_pp3}) with greatly enhanced accuracy. To derive $\epsB$ we calculate both the magnetic-field and total energy densities in the flow-comoving frame, and then average the result over the specified time interval of the steady-state solution. We identify three regions of qualitatively different behavior.
In the first, far upstream region, the magnetization rises approximately as power law in distance from the shock front, and the magnetic field growth is likely due to relativistic filamentation (Weibel) instability. The power-law index slightly varies --- we observe $\epsB \propto x^{-1.2}$ in the \simmain{} simulation and $\epsB \propto x^{-1}$ in the \simlong{} simulation.

In the second region, which comprises the shock front and the immediately adjacent upstream, the magnetization apparently follows compression law, $\epsB \propto u_\mathrm{_{EM}}^{-2}$, as suggested by \cite{Lemoine2019}, where $u_\mathrm{_{EM}} = \beta_\mathrm{_{EM}} \Gamma_\mathrm{_{EM}}$ is the 4-velocity corresponding to the ``electromagnetic velocity'' calculated as $\beta_\mathrm{_{EM}} = \sqrt{\langle E_y^2 \rangle /\langle B_z^2 \rangle}$.
Here the magnetic field grows as if it were frozen-in into a fluid moving at $\beta_\mathrm{_{EM}}$. 
In the third, downstream region, the magnetic field decays also as power law in the distance from the shock front, $\epsB \propto |x|^{-0.8}$. All simulations exhibit the same downstream decay law except \simcold{}, where the power law index is $\epsB \propto |x|^{-1}$. 
This difference is likely not due to the lower upstream temperature itself, but due to the low density contrast that prevents the formation of slowly decaying nonlinear magnetic field structures.

The local properties of the magnetic turbulence also change as a fluid element passes from the upstream region to the downstream region through the shock front.
The lower panels (A-F) of Fig.~\ref{fig_pp3} show the time-averaged spatial power spectra at different locations with respect to the shock front. Close to the upstream boundary (panel A), the spectrum is highly anisotropic and dominated by modes with $k_y\gg k_x$, which are typical of filamentation instability. Then (see panels B-C), the modes with $k_y\sim k_x$ begin to appear in the power spectra. They grow in amplitude as a fluid element moves towards the front and become dominant close to the front. 
The power spectrum remains anisotropic until then. The front passage greatly perturbs the power spectrum (panel D), which becomes ring-shaped like in the downstream (panels E-F), where it retains that ring-like shape. 
Further downstream, the evolution of the magnetic turbulence spectra is consistent with the gradual decay of the existing modes (the higher-$k$ modes decay faster, as expected). Although some of the lower-$k$ modes do show growth, their growth rates are too small to claim that the magnetic energy is transferred from smaller to larger spatial scales.

\begin{figure*}
   \centering
   \includegraphics[width=0.95\textwidth]{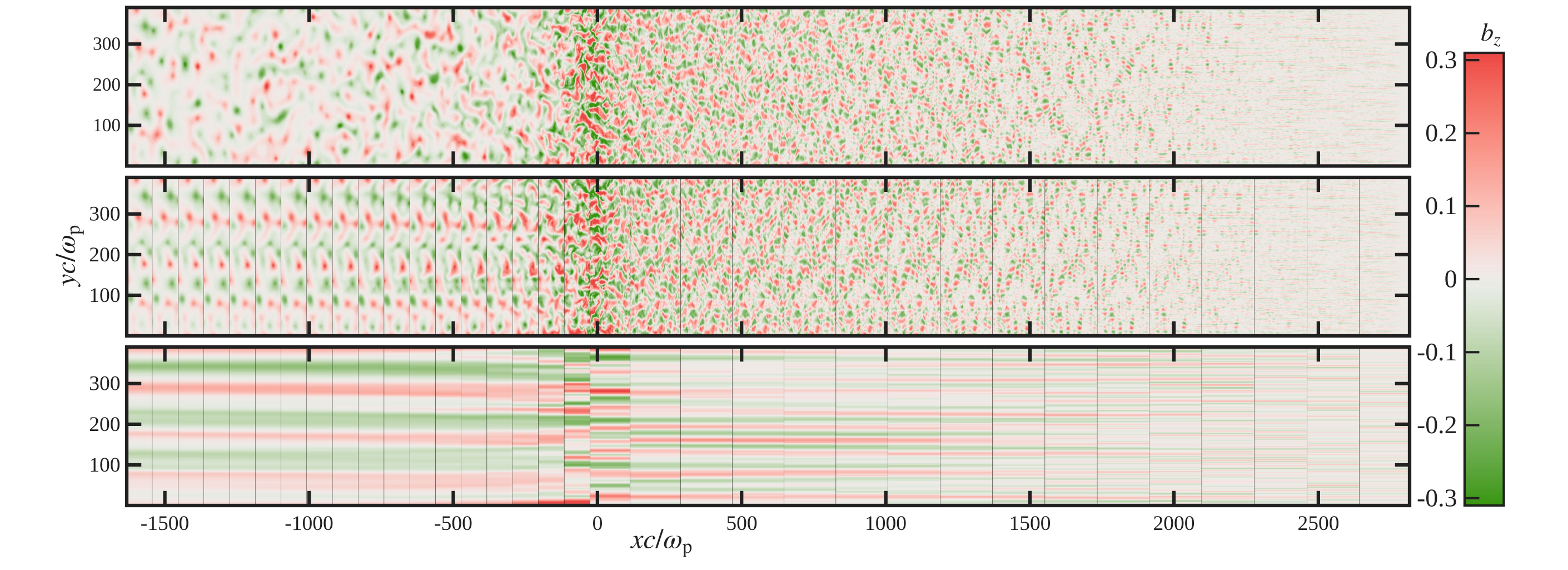}

   \caption{
   Normalized out-of-plane magnetic field $b_z = B_z/\sqrt{8\pi n_{0}E_0}.$
   {\it Upper panel}: instantaneous snapshot at $t\approx 96600\omega_\mathrm{p}^{-1}$ after the start of the simulation.
   {\it Middle panel}: composite map showing the evolution of the magnetic field within a fluid element in the locally-comoving (Lagrangian-like) frame. The element enters the simulation domain at $t = 91100\,\omega_p^{-1}$ (the rightmost snapshot) and approaches the downstream boundary at $t \approx 96600\,\omega_p^{-1}$ (the leftmost snapshot). 
   Snapshots are taken every $\delta t=160\omega_p^{-1}$. 
   For each snapshot the positions of the element’s leading and trailing edges are calculated by integrating the bulk velocity along the flow. 
   More details on the technique are given in the text.
   The vertical lines mark the boundaries between consecutive snapshots.
  {\it Bottom panel}: each segment of the middle panel is averaged over the 
  $x$-coordinate and multiplied by 2. This way we expose the secular evolution of the magnetic field topology. 
  }
   \label{comb_field}   
\end{figure*}

\begin{figure}
   \centering
   \includegraphics[width=1.0\columnwidth]{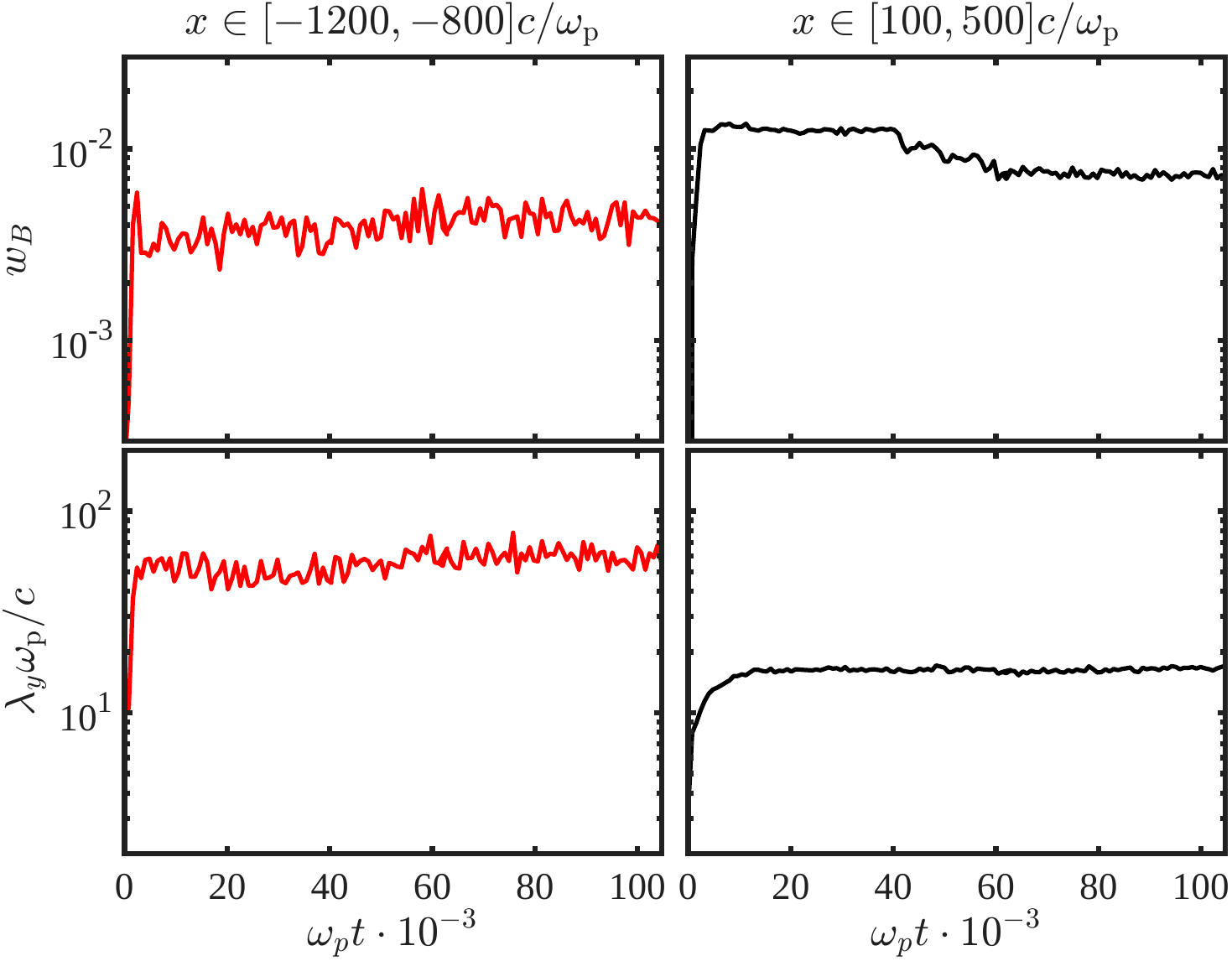}   
   \caption{Evolution of magnetic field characteristics behind (left {panels}) and ahead of (right {panels}) the shock front in \simmain{} simulation. 
   {\it Top panels:} the average magnetic energy density in  fixed downstream ($x \in [-1200, -800]c/\omega_\mathrm{p}$) and upstream ($x \in [100, 500]c/\omega_\mathrm{p}$) slabs. 
   {\it Bottom panels:} the magnetic-field transverse coherence scale $\lambda_y$. All quantities are given as functions of simulation time. }
   \label{fig_4}
\end{figure}

Another perspective of magnetic-field evolution along the flow  is given in Fig.~\ref{comb_field}. 
It corresponds to the steady-state interval of the \simmain{} 
simulation, at time $t \simeq 10^5\,\omega_\mathrm{p}^{-1}$ from the start.
The upper panel shows an instantaneous snapshot of the magnetic field. In the upstream, the overall picture is similar to what was previously observed in numerous simulations in the downstream-comoving setup: the magnetic field is generated in the furthest upstream being shaped as long filamentary structures (due to the Weibel instability), then these filaments rearrange themselves into net-like structures (forming low-density cavities inside highly-magnetized boundaries), which grow in amplitude until reaching the front, where the magnetic structures get stirred.  We do not observe a global asymmetry in the polarities of upstream magnetic structures and in the currents of returning particles, reported in \citep{Groselj2024}. Instead, polarity of every cavity is seemingly random and independent of the neighboring cavities. The pattern of these magnetic structures also differs from the observed in \citep{Groselj2024}: their boundaries are not as thin and localized, and the density contrast is relatively smaller.
All the smaller-scale structures disappear shortly after the front passage, whereas prominent larger-size magnetic spots quickly acquire round shape and, slowly evolving, persist until reaching the downstream boundary. The magnetic field distribution in an individual soliton-like spot has approximately Lorentzian shape. As the spots evolve, they shrink in radius whereas their amplitudes barely change. Combined, these effects lead to decreasing average magnetization as the distance to the shock increases.

The middle panel of Fig.~\ref{comb_field} presents a Lagrangian view of the magnetic‑field evolution. A fluid element that enters at the upstream boundary is tracked as it is advected through the simulation box. 
We store full‑domain snapshots of $b_z$ at intervals $\delta t = 160 \omega_p^{-1}$ and, integrating the local bulk flow velocity, determine the instantaneous position $x(t)$ of the plasma element’s leading edge. For the first snapshot we take the part of the domain from the injection (upstream) boundary to $x(t_1)$; for each subsequent snapshot we retain only the newly traversed interval between $x(t_{i+1})$ and $x(t_i)$. These pieces are placed side‑by‑side (separated by thin vertical lines), giving a time-lapse sequence of the magnetic-field evolution.
 One may see the fluid element shrinking as it is compressed in the decelerating flow. (The upper panel, being a global snapshot at the end of the interval, allows one to see the final position of the tracked element in the context of the entire simulation domain.) 
Following a fluid element, we find that away from the front the magnetic-field evolution is dominated by independent size and amplitude changes of these spots, without notable reconnections or coalescence.  
This morphology of the downstream magnetic field is the same in all our simulations. 
In our simulations, the downstream region has fixed size and we can not track these spots indefinitely far from the front. 
Nevertheless, our simulations suggest that higher-field spots decay slower compared to lower-field spots of the same size.

The bottom panel shows the same data as in the middle panel after each segment has been averaged over the $x$-axis. Together, the two panels demonstrate that the magnetic-field topology changes only in a thin region at the shock front; away from the front, both upstream and downstream, the evolution is chiefly a decrease in amplitude while the field topology remains essentially unchanged.

Figure \ref{fig_4} shows the time evolution of the magnetic-field energy density $w_\mathrm{_B}$ and the transverse correlation length of the magnetic field $\lambda_y$ within fixed {regions} downstream (left panels) and upstream (right panels) of the shock (see figure caption for exact size and location). The evolution patterns of those quantities in our simulations are clear:  both $w_\mathrm{B}$ and $\lambda_y$ experience short initial rise and then become asymptotically constant in both upstream and downstream. No significant evolution is observed at later time $t>25000\omega_p^{-1}$.
In contrast, \citet{Groselj2024} reported only the general trend for increasing correlation length on the full time span of their simulation, although the short-time variations are large.

\begin{figure}
   \centering
   \includegraphics[width=1.0\columnwidth]{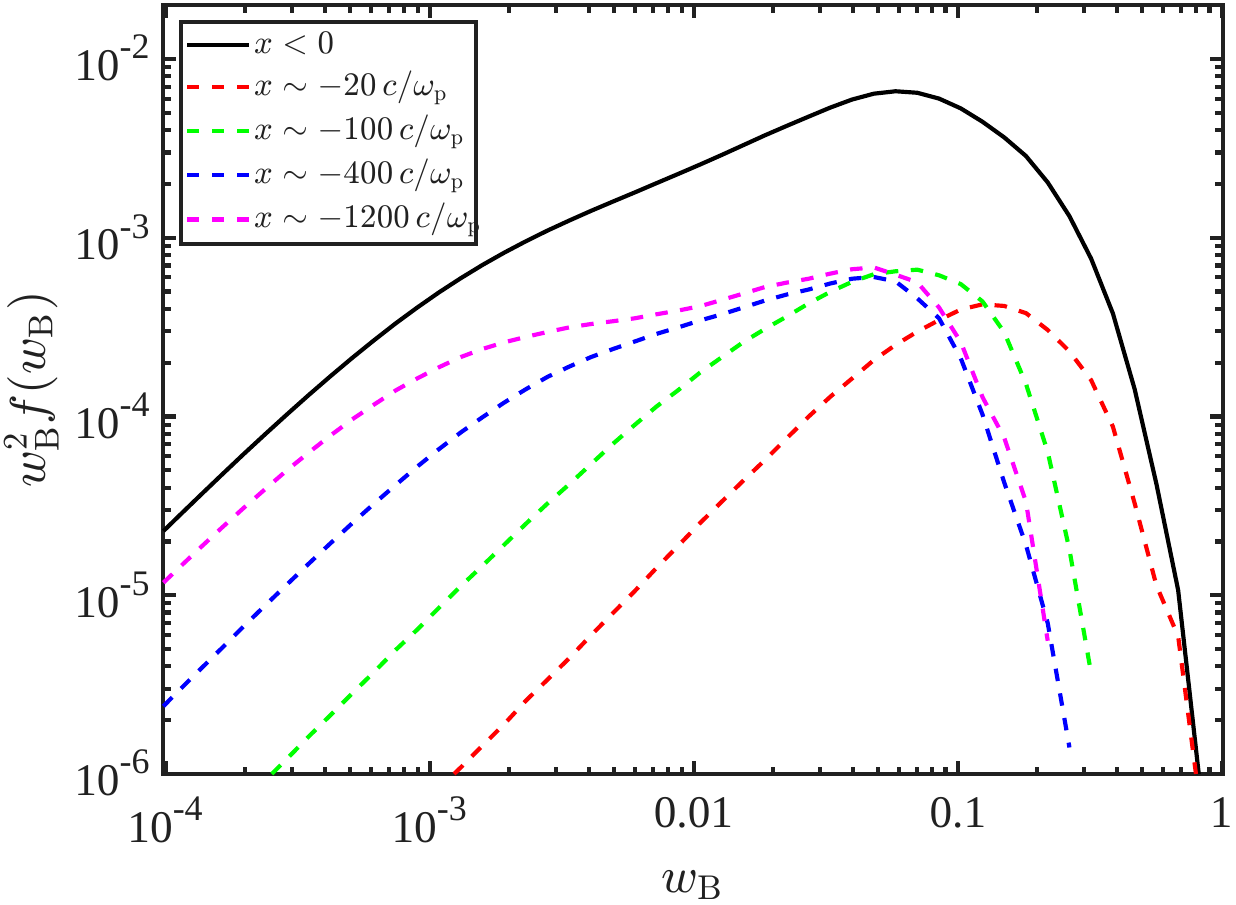}   
   \caption{ 
   The distribution of magnetic energy per logarithmic interval in $w_\mathrm{_B}$ (here $f(w_\mathrm{_B})$ is the probability distribution over $w_\mathrm{_B}$), averaged over the steady-state interval of the \simmain{} simulation.
   The distribution for the entire downstream region (solid black line) is the envelope of local distributions at different distances from the front (dashed colored lines).   
   }  
   \label{fig_10X}
\end{figure}     

Figure~\ref{fig_10X} 
illustrates the distribution of magnetic energy across regions with different magnetic field strengths, both throughout the entire downstream and within thin slices at specific distances from the shock front.
While the magnetic energy is concentrated in regions with $w_\mathrm{_B} \gtrsim 0.05$ (the peak of the $w_\mathrm{_B}^2 f(w_\mathrm{_B})$ curve),  most of the volume has much lower magnetic energy density $w_\mathrm{_B} \lesssim 2\times 10^{-3}$ (the break of the curve), with this discrepancy growing further downstream.
Furthermore, the magnetic energy density of the energy-carrying regions decreases only by a factor of $\approx 2$ when going from $x=-20\ c/\omega_p$ to $x=-1200\ c/\omega_p$, whereas the average value decreases by a factor of $\approx 20$. We conclude that the magnetic spots are becoming smaller in size, but their amplitude, as long as we can track them, remains almost constant.
Both locally and when integrated over the entire downstream region, the distribution of magnetic energy density in our simulations is significantly skewed toward higher values. We find a power-law asymptote $w_\mathrm{B}^2f(w_\mathrm{B})\propto w_\mathrm{B}^{0.65}$ for the full downstream length, with a tendency toward a steeper slope when shorter segments of the downstream are considered. 
Over this observable lifespan, the magnetic field distribution around the centers of these soliton-like magnetic spots is approximately Lorentzian, which corresponds to $w_\mathrm{B}^2f(w_\mathrm{_B})\propto w_\mathrm{}^{1/2}$ asymptotical behavior, consistent with the suggestion by \citet{Groselj2024}.

Table~\ref{SummaryTable2} summarizes the steady-state characteristics measured in all our simulations.
For the moderate-temperature setup ($T_\mathrm{up} = mc^2$), variations in numerical realization produce only minor changes (except for the downstream transverse coherence scale in the \simnarrow{} simulation, which is clearly underresolved in the y-direction) in the downstream coherence length, density contrast, particle spectrum and magnetization, indicating that the downstream steady state is robust. 
The \simlong{} simulation shows a larger $\lambda_{y}^\mathrm{(up)}$ and a substantially longer upstream settling time, consistent with slower convergence of the far-upstream structure. Changing the upstream temperature produces the strongest trend: the \simcold{} simulation has systematically smaller coherence scales, lower magnetization and weaker density contrast, while the \simhot{} simulation shows the opposite behavior. 

\squeezetable
\begin{table*}
\caption{Steady-state {characteristics} for the simulations listed in Table~\ref{SummaryTable}. 
The transverse correlation lengths (coherence scales) of the magnetic field, $\lambda_{y}^\mathrm{(ds)}$ and $\lambda_{y}^\mathrm{(up)}$, are measured, respectively, in the downstream slab $x\in(-700, -500)c/\omega_p$ and the upstream slab $x\in(500, 700)c/\omega_\mathrm{p}$ (coordinates are relative to the front position). The steady-state settling time $t_\mathrm{s}$ is obtained by fitting the time evolution of residuals of the particle spectra with an approach explained in the text, the particle spectra in the slab $x\in(-700, 700)c/\omega_\mathrm{p}$ were used. 
$\epsilon_\mathrm{B}^\mathrm{(front)}$ is the magnetization in the near-front region $x\in(-20, 5)c/\omega_p$. Density contrast and Lorentz factor  $\gamma_\mathrm{acc}^\mathrm{(front)}$(which is the second maximum of $\gamma^3f(\gamma)$ function) both are measured in the same near-front region. All the values are averaged over the steady-state time interval.
}

\begin{ruledtabular}
\begin{tabular}{p{4cm}p{1.6cm}p{1.6cm}p{1.6cm}p{1.6cm}p{1.6cm}p{1cm}}
Simulation name & $\lambda_{y}^\mathrm{(ds)}\omega_p/c$ & $\lambda_{y}^ \mathrm{(up)}\omega_p/c$ & $t_\mathrm{s}\omega_p$ &  $\epsilon_\mathrm{B}^\mathrm{(front)}$ & density contrast & $\gamma_\mathrm{acc}^\mathrm{(front)}$   \\    \hline
\simmain & $60$ & $16$  & $4000$ & $2.7\cdot10^{-2}$ & $3.5$ & $960$ \\ 
\simshort & $60$ & $17$  & $2800$ & $2.8\cdot10^{-2}$ & $3.5$ & $980$ \\ 
\simlong & $56$ & $23$  & $7900$ & $2.1\cdot10^{-2}$ & $2.9$ & $1200$ \\ 
\simnarrow & $40$ & $15$  & $4000$ & $2.7\cdot10^{-2}$ & $3.5$ & $1000$ \\ 
\simwide & $63$ & $17$ & $4200$ & $2.6\cdot10^{-2}$ & $2.8$ & $980$ \\ 
\simcold & $33$ & $11$ & $3800$ & $1.2\cdot10^{-2}$ & $1.9$ & $240$ \\ 
\simhot & $79$ & $23$ & $4300$ & $4.0\cdot10^{-2}$ & $6.6$ & $3870$ \\
\simshortl & $51$ & $17$ & $3800$  &  $2.5\cdot10^{-2}$ & $2.9$ & $1000$
\end{tabular}
\end{ruledtabular}
\label{SummaryTable2}
\end{table*}

To check the possible effect of downstream and upstream sizes, we run several simulations with different lengths. We find that the simulation with longer upstream has larger correlation length upstream of the shock front.
However, $\lambda_y$ remains unchanged downstream of the shock front, no matter whether we change the upstream length, the downstream length, the spatial resolution, or increase the width of the simulation box. 
The same is true for other quantities that we track, including density contrast, magnetization, etc. -- while the stationary state in the upstream may change a little from simulation to simulation, the downstream quantities are very stable. One exception is the \simnarrow{} simulation, which exhibits a shorter correlation length in the downstream region compared to the \simmain{} run. 
This indicates that a reduced transverse box size cannot fully capture long-wavelength modes. {However, increasing the transverse width from \(L_y=400c/\omega_p\) in \simmain{} to \(L_y=1000c/\omega_p\) in the \simwide{} run, gives a downstream coherence scale of approximately \(63c/\omega_p\), close to the \simmain{} result. The characteristic particle energy is also nearly unchanged (\(\gamma_\mathrm{acc}\simeq960\) versus 980). This demonstrates that the width of the \simmain{} run is sufficient.} Also, the choice of a specific numerical setup has a minor effect on the resulting steady-state particle distribution.

{ The persistent slow evolution of the far upstream does not, by itself, indicate that an asymptotic steady state for the downstream cannot be reached. Even the absence of the true steady state upstream would not have significant consequences in practice, because it is the downstream that determines the observational appearance of a shock.} 

\begin{figure}
   \centering
   \includegraphics[width=1.0\columnwidth]{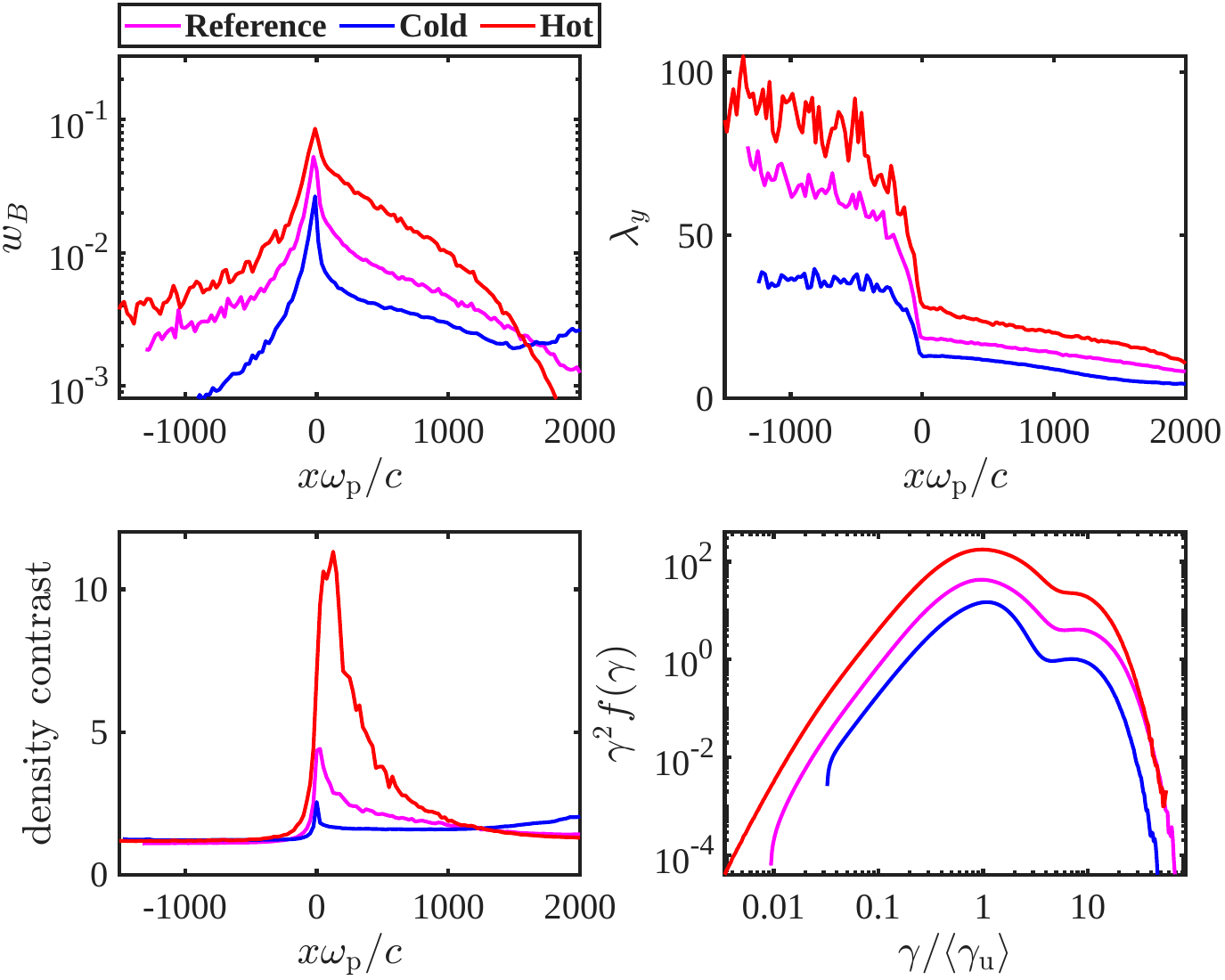}   
   \caption{Comparison of the steady state for magnetic energy density, transverse correlation length, density contrast as functions of coordinate along the flow and particle energy distribution in the \simmain, \simhot, and \simcold{} simulations. Each curve is time averaged over the interval of the steady-state behavior, $t\in[76000, 105000]\omega_\mathrm{p}^{-1}$ for \simmain{}, $t\in[60000, 81000]\omega_\mathrm{p}^{-1}$ for \simcold{} and $t\in[50000, 62000]\omega_\mathrm{p}^{-1}$ for \simhot{}. The normalization constant $\langle \gamma_\mathrm{u} \rangle$ in the particle distribution function is the average {Lorentz factor} of the upstream particles at the moment of their injection at the boundary, 
   $\langle \gamma_\mathrm{u} \rangle \equiv E_0 / mc^2$. The upstream temperatures that we used, $T_\mathrm{up} = \{0.1, 1, 5\} mc^2$, correspond to the values $\langle \gamma_\mathrm{u} \rangle \approx \{31, 108, 500\}$, respectively. 
   }
   \label{fig_n4}
\end{figure}

To explore how the upstream temperature influences the the shock's properties, we perform three simulations with the upstream temperature at injection boundary $T_\mathrm{up} = \{0.1, 1, 5\}mc^2$. All of them reached the steady state in the sense described in the previous section. Plots of the magnetic energy density, the correlation length, the density contrast, and the particle distribution functions are presented in Figure \ref{fig_n4}.
Increasing the upstream temperature from $T_{up} = 0.1\,m_ec^2$ up to $T_{up} = 5\,m_ec^2$ raises the magnetic energy density, increases the upstream and downstream coherence lengths, and shifts  the characteristic energy of accelerated particles to higher Lorentz factors. 

Also, we observed a trend of decreasing density contrast with lower upstream temperature, which seems to contradict the results of \citep{Groselj2024}, whose simulation at $T \approx 0$ yields a significantly larger density contrast than we measure at $T = 0.1\,mc^2$. 
This discrepancy may stem from differences in the simulation setups. 
In our setup, the incoming upstream flow {encounters} a finite flux of outgoing accelerated particles immediately at the boundary, triggering a rapid onset of the filamentation instability.
Conversely, in the standard setup, even the very first outgoing particles do not reach the upstream boundary, 
and the filamentation instability develops gradually, with its growth rate starting from zero and slowly increasing over time.
Note that it is unclear which setup better matches the real physical situation, where even a very small magnetic field frozen into the upstream plasma would limit the precursor's extent.

{A qualitatively similar temperature dependence has recently been reported by \citet{Kamiido2024}. Their relativistic-shock simulations show a larger density fluctuation
contrast for relativistically hot upstream plasma compared with a cold upstream. }

All particles in the simulation box can be divided into two components: the bulk flow particles (injected at the upstream boundary and moving downstream on average with the flow's velocity) and the accelerated particles (those that were reflected from the front or spent a long time there). They are separated by comparing the individual particle lifetime to the time taken for the bulk flow to reach the particle's location.
In the upstream region, the two populations are well-separated. Formally, we define accelerated particles as those with a lifetime exceeding the time it takes to travel from the upstream boundary to the shock front and then back to the particle's location with the speed of light, i.e., \(t_\mathrm{life} > (L_\mathrm{up}/c) (1+x/L_\mathrm{up})\). In the steady state of the \simmain{} simulation, the accelerated particles move upstream with an average speed of $\beta\approx 0.3$ relative to the front. 
Although we have not directly measured this speed in other simulations, the overall similarity in their distributions of accelerated particles suggests that it does not change significantly.

In the downstream, the division is to some extent arbitrary. Here, we define accelerated particles by the condition \(t_\mathrm{life} > L_\mathrm{up}/c + 4|x|/c\),
which selects particles that spend in the downstream twice as much time as the bulk flow at a given point. In the downstream region, the energy fraction in accelerated particles is about 15\%.

\begin{figure*}
   \centering
   \includegraphics[width=1.0\textwidth]{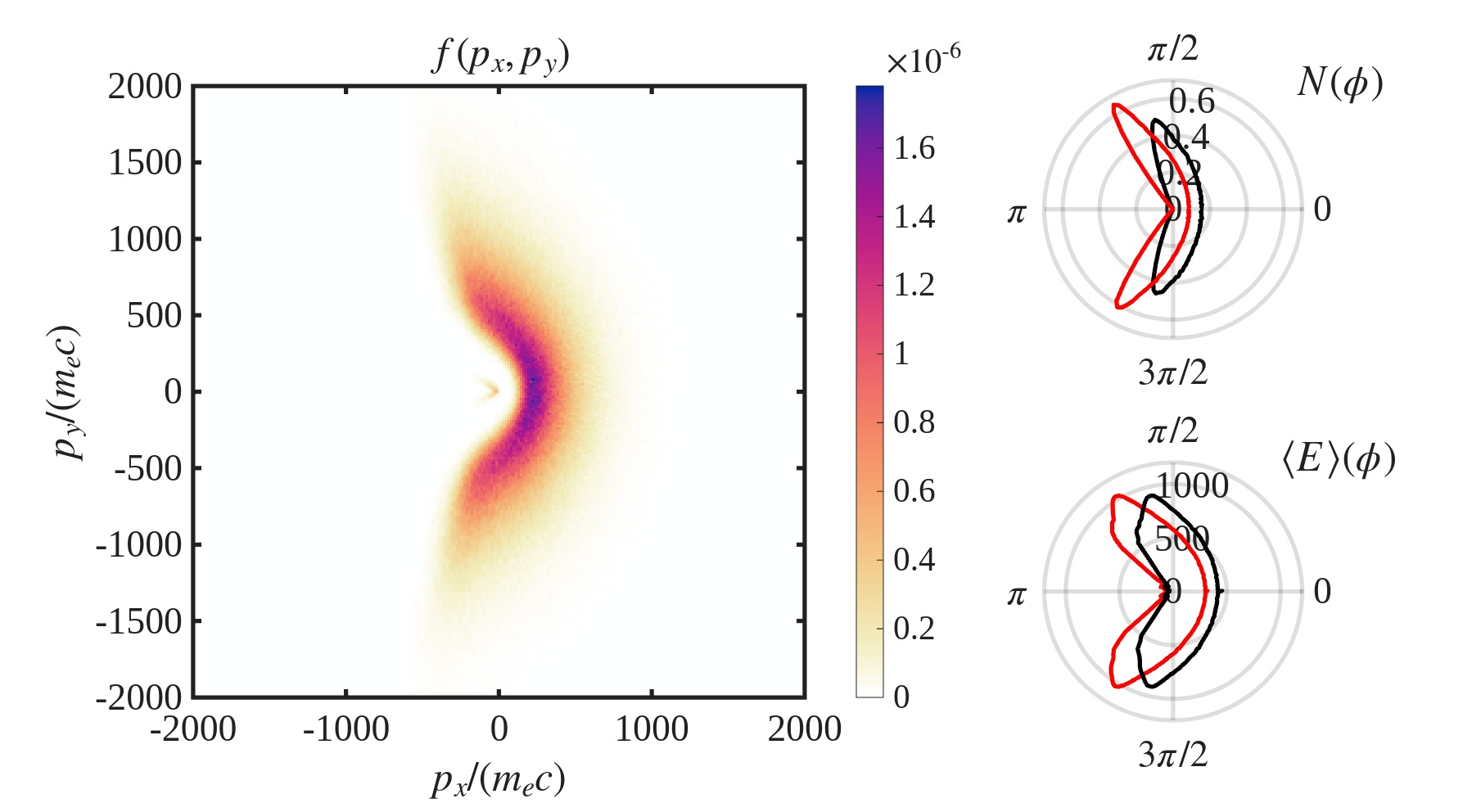}
   \caption{The distribution of accelerated particles (see text for the definition) in the upstream region \(x\in[1400, 1800]\,c/\omega_p\) for the \simmain{} simulation at $t = 105000\,\omega_p^{-1}$.  
   {\it Left panel:} distribution of accelerated particles in the momentum space (front-comoving frame). The normalization is $\int f(p_x, p_y) (dp_xdp_y)/(m_e c)^2 = 1$.
   {\it Right panels:} angular dependences of the accelerated particle count (top) and their average energy (bottom).
   Angles are measured from the downstream direction.
   The normalization of the particle count is $\int f(\phi) d\phi = 1$.
   The black curves correspond to the values measured in the simulation (i.e., front-comoving) frame, and the red curves are for the values that are evaluated in a reference frame moving upstream relative to the simulation frame at a velocity \(\beta \approx 0.31\) that matches the average velocity of the accelerated particles.
   }
      \label{fig_ap}
\end{figure*}  

The distribution of accelerated particles is shown in Fig.~\ref{fig_ap}. This distribution is highly anisotropic. The most energetic particles surf the shock front instead of bouncing between the upstream and the downstream, contrary to expectations for Fermi acceleration. We consider this as an indication of weak scattering --- to scatter efficiently, the high-energy particles must move almost along the shock plane, and they are unlikely to return after leaving the region of strong magnetic turbulence. The transition to the weak-scattering regime is presumably the primary factor limiting the maximum particle energy.

\begin{figure}
   \centering
   \includegraphics[width=1.0\columnwidth]{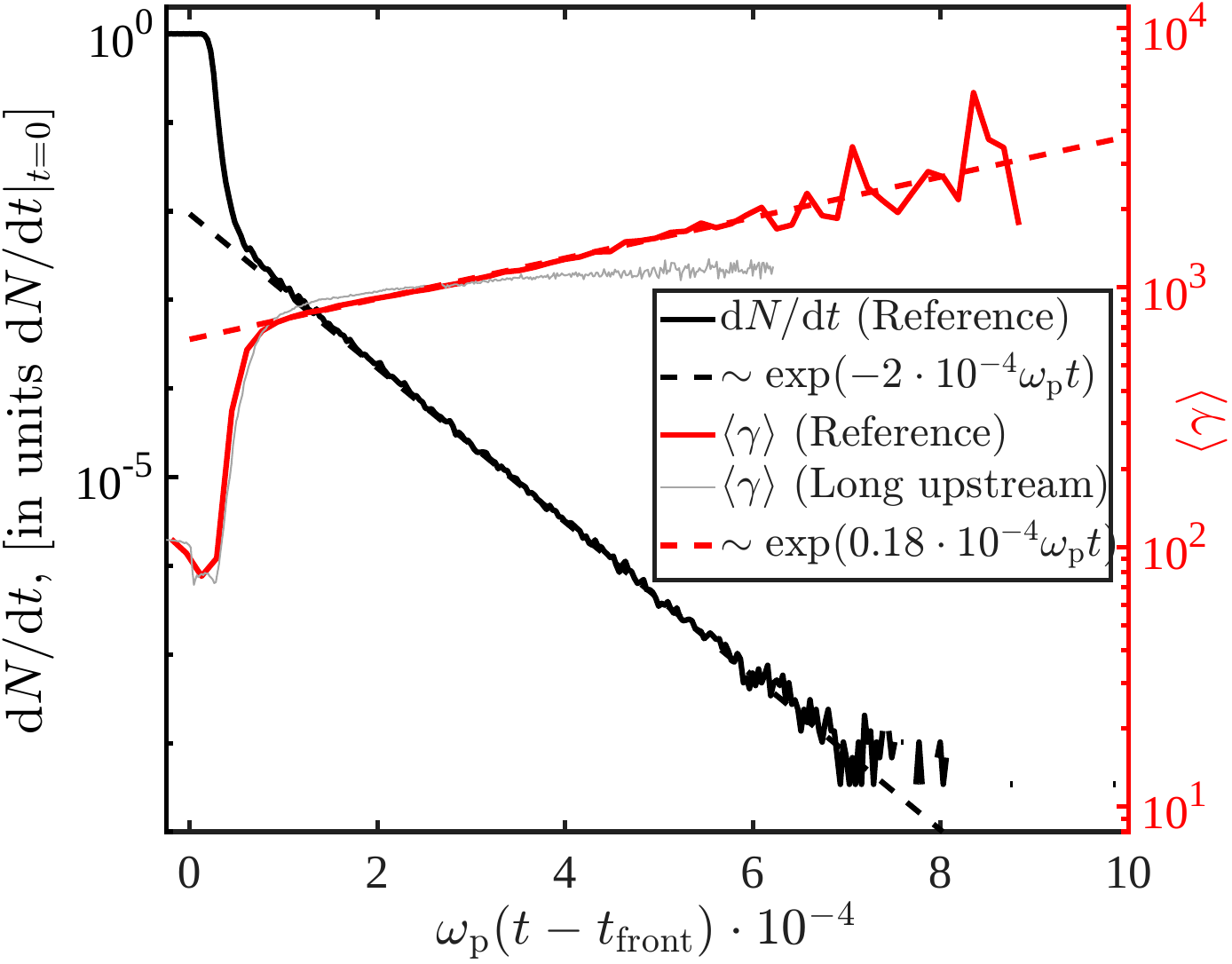}   
   \caption{ Normalized number of particles per unit lifetime (black curve, axis on the left) and their average Lorentz factor {$\langle \gamma \rangle$} (red curve, axis on the right) as functions of the time that particles have spent in the simulation box. The data correspond to the end of \simmain{} simulation. The dashed red lines show asymptotic behavior. {For comparison, the thin gray line shows {$\langle \gamma \rangle$} for the \simlong{} simulation. In order to compare different simulations properly, we subtract from the particles' lifetime the time needed for the flow to reach the shock front $t_\mathrm{front}$. }   
   }
   \label{fig_11}
\end{figure}

\begin{figure}
   \centering
   \includegraphics[width=1.0\columnwidth]{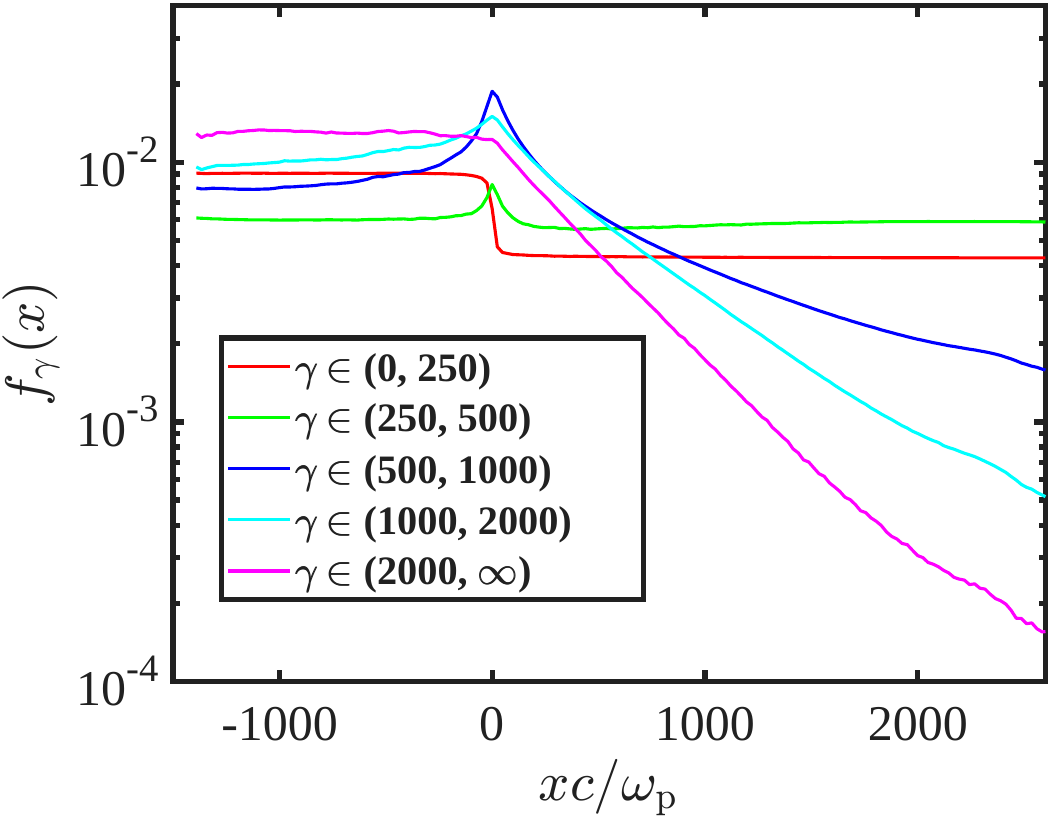}   
   \caption{{Spatial distributions of particles with different Lorentz-factors across the shock. Each curve represents the particle distribution \(f_\gamma(x)\), normalized as \(\int f_\gamma(x)\,dx=1\). The Lorentz-factor intervals are \(\gamma\in(0,250)\) (red), \(\gamma\in(250,500)\) (green), \(\gamma\in(500,1000)\) (blue), \(\gamma\in(1000,2000)\) (cyan), and \(\gamma\in(2000,\infty)\) (magenta). }
   }
   \label{fig_a1}
\end{figure}

Figure~\ref{fig_11} shows the distribution of particles over their lifetime in the simulation box  and the average Lorentz factor as a function of particles' lifetime, both for the \simmain{} run. 
The fast acceleration stage apparently concludes near $t \approx 12000\omega_\mathrm{p}^{-1}$, beyond which the acceleration rate diminishes. However, the long-time limit still exhibits an exponential growth of the average energy with particle lifetime. Concurrently, the number of particles remaining within the simulation box decreases over time, also following an exponential asymptote. Together, these two asymptotic behaviors should formally produce a power-law distribution $f(\gamma) \propto \gamma^{-11}$, which is extremely soft and contains a vanishing number of particles. 
Moreover, such an exponential increase in average particle energy with time is difficult to reconcile with Fermi acceleration, which normally yields a linear energy growth. Further investigation reveals that this increase in average energy does not stem from actual acceleration. Instead, it is a selection effect: lower-energy particles escape the simulation box faster than higher-energy ones, whose trajectories are aligned with the shock plane, resulting in a small $x$-component of their velocity. 
{A similar curve plotted for the \simlong{} run clearly demonstrates that acceleration stalls at the same moment, indicating that the duration of the active acceleration phase is independent of the upstream length.}

{We now return to the settling time $t_\mathrm{s}$ in Table~\ref{SummaryTable2}. It clearly depends on the simulation-domain length, apparently reflecting initial-condition transients rather than the time needed to establish the acceleration region. Although we cannot robustly quantify the true shock settling time, the simulations show that the region around the shock front reaches a steady state well before the measured $t_\mathrm{s}$. This delay mainly reflects the time needed for accelerated particles to fill the entire domain: since the most energetic particles propagate almost along the front, they reach the boundaries slowly. Hence, even after the shock-front structure has largely ceased to evolve, the measured settling time still depends on box length, because full-distribution equilibration cannot be readily decoupled from shock evolution.}

{Figure~\ref{fig_a1} demonstrates the spatial distributions of particles in different Lorentz-factor intervals, revealing qualitative changes with energy. The low-energy population (red curve) follows the hydrodynamic structure of the flow, with the density jump at the front. In contrast, particles with intermediate energies (green, blue, and cyan curves) reach maximum density near the front, suggesting that 
they originate there.
This is a clear signature of the acceleration process being localized to region near the shock front. The highest-energy particles (magenta curve), however, do not exhibit a maximum at the front. This means that the particles with higher energies are not accelerated efficiently in near-front region. }

{The upstream distribution of the higher-energy particles is more concentrated towards the front. The density of the highest-energy particles at the upstream boundary drops by two orders of magnitude relative to their density at the front. Thus, if an increase in the upstream length does affect the escape of accelerated particles, it primarily concerns low-energy particles. The effect on the highest-energy particles is very limited. }

\section{Summary}
\label{sec:summary}

We introduced a new front-comoving PIC setup designed to follow the evolution of collisionless shocks for extremely long times. We ran several simulations for the relativistic pair shocks up to times $\gtrsim 10^{5}\,\omega_p^{-1}$. The key technical element is a moving-wall downstream boundary condition combined with continuous upstream injection, which enables a fixed-size computational domain and makes it feasible to reach durations well beyond standard downstream-comoving simulations. 
At the same time, several limitations should be kept in mind. The front-comoving setup necessarily introduces some constraints: the accelerated-particle flux hits the upstream right at the border of the simulation box, in contrast with slowly raising flux of reflected particles in the simulations with effectively infinite upstream; the moving-wall boundary effectively reflects all the particles at a fixed distance downstream of the front. While both effects can be controlled by varying the box dimensions (and were tested here through several realizations), they remain potential sources of bias for the detailed far-upstream precursor structure.

The results of our simulations independently verify several previously reported findings, thereby demonstrating their robustness against variations in numerical setup. Specifically, we observe that the magnetic field has a filament-like structure far upstream, changing to cavity-like structures closer to the front. Associated cavities also form in the particles' density.
The average magnetization peaks at the shock front at a level of $\epsB \sim 0.03$ ($w_B\sim 0.07$). In the immediate vicinity of the front, the magnetic field strength follows the "compression law": it grows as if it were frozen into the fluid and advected with 
the velocity $\beta_\mathrm{EM}$, which in this region only slightly deviates from the bulk hydrodynamic velocity. The particle distribution is non-thermal, and there is a population of particles with energies much larger than the equipartition energy.

At the same time, some details of our results differ from what was previously reported in downstream-comoving PIC simulations. We have never observed the globally coherent polarization of the magnetic field near the shock front that was reported by \cite{Groselj2024}. 
Density and magnetization contrasts in the vicinity of the shock front are substantially smaller in our simulations, though they increase towards higher upstream temperature, reaching
a visually similar amplitude for the \simhot{} simulation ($T_\mathrm{up} = 5 m_e c^2$).
The magnetic field in the downstream is highly non-uniform: most prominent of the small magnetic spots have a magnetization above $0.2$ in the \simmain{} simulation (this is a factor of $\approx 4$ higher than reported by \cite{Groselj2024}), while most of the volume is weakly magnetized, with the magnetization being much lower than the average value.
Determining the energy content of non-thermal particles carries an intrinsic uncertainty in 2D3V simulations: particles with different momentum components transverse to the simulation plane ($p_{z}$) may thermalize differently, and the standard Maxwell–J\"uttner distribution does not form. With these necessary caveats in mind, we estimate that the fraction of energy transferred to the non-thermal component rises with increasing upstream temperature, from $\sim 0.1$ for $T_\mathrm{up} = 0.1$ to $\sim 0.25$ for  $T_\mathrm{up} = 5$. This differs from the results of \cite{Groselj2024}, who reported a still larger fraction for $T_\mathrm{up} = 0$.
If accelerated particles are defined in a more robust way through their lifetime in the simulation box, then the energy fraction in these particles in the downstream is $\approx 0.15$ for the \simmain{} simulation.

The principal new finding of our record-long runs is a clear evidence that the system evolves toward a steady-state for the quantities most relevant to radiation modeling.
The evolution of the shock parameters (correlation length, magnetization, maximum energy of accelerated particles, etc.) ceases and the system reaches a late-time stationary state. 
Importantly, the time required to reach this steady state is shorter in the simulation with shorter downstream region, whereas the parameters of the steady-state downstream exhibit no dependence on the downstream or upstream lengths. {Consequently, our simulations provide evidence for a late-time downstream steady  state. At the same time our results do not exclude a possibility that far upstream of a shock with exactly zero magnetic field may not reach a steady state.}

Next, we find that the upstream magnetization increases as a power-law with distance as $\epsB \propto |x|^{-1.2}$. 
In the downstream region, the magnetization also decreases as a power-law of distance to the shock front, $\epsB \propto |x|^{-0.8}$.
The topology of the magnetic field changes dramatically during the transition from the highly magnetized region near the shock front into the far downstream, where the field decomposes into a set of isolated and independently evolving magnetic spots. Closer to the shock, the local magnetization follows a Gaussian-like distribution, while further downstream, this distribution becomes consistent with that expected from a set of independent spots with a Lorentzian magnetization profile.

Finally, we find that Fermi-type acceleration of particles stalls upon reaching an energy about an order of magnitude larger than the equipartition energy. The distribution of accelerated non-thermal particles is highly anisotropic, with the highest-energy particles ``surfing'' the shock front and moving almost parallel to it.
This phenomenon is a potential source of net polarization in the synchrotron emission from relativistic shocks.

{We emphasize that the front-comoving and downstream-comoving approaches have different numerical advantages and limitations and should be viewed as complementary.}
The key drawback of the downstream-comoving setup is the quadratic increase of the simulation cost with physical time.
On the other hand, this setup is less prone to numerical instabilities in the upstream, that allows one to model a zero upstream temperature (keeping in mind all necessary caveats regarding upstream heating by $e^{-}e^{+}$-pairs produced by radiation from the shock). 
Conversely, the front-comoving setup features a lower simulation cost due to a fixed domain size, with the computational cost only linearly growing with physical time, thus allowing one to simulate shock evolution over much longer timescales.
On the other hand, a low-temperature upstream in this setup is heavily influenced by unphysical electromagnetic modes {(due to the development of NCI)}, effectively setting a lower limit on the upstream temperature that can be reliably modeled (again, this may not be a limitation in practice).
While neither setup exactly corresponds to a realistic shock model, both can capture the essential physics.
The long-duration capability of the front-comoving setup opens a potential way to systematically study shock microphysics on the radiative-loss time scales, including the self-consistent evolution of particle spectra and magnetic field decay.

\begin{acknowledgments}
ED acknowledges support from the Russian Science Foundation (grant no. 24-12-00457). 
The authors would like to thank D.~Gro{\v{s}}elj for critically reading the manuscript and highlighting the points that required further clarification.
\end{acknowledgments}

\appendix
\section{The moving-wall boundary condition}
\label{sec:bc}

We suggest a boundary condition that can be used in a situation where there is continuous flow of particles through the simulation-box boundary where a periodic  boundary condition is inappropriate (for example, the hydrodynamic velocity differs at opposing boundaries). One obvious application of this boundary condition is simulations of shocks in the front-comoving frame. Here we describe two statistically equivalent implementations of the moving-wall boundary condition.

\begin{figure}
   \centering
   \includegraphics[width=0.64\columnwidth]{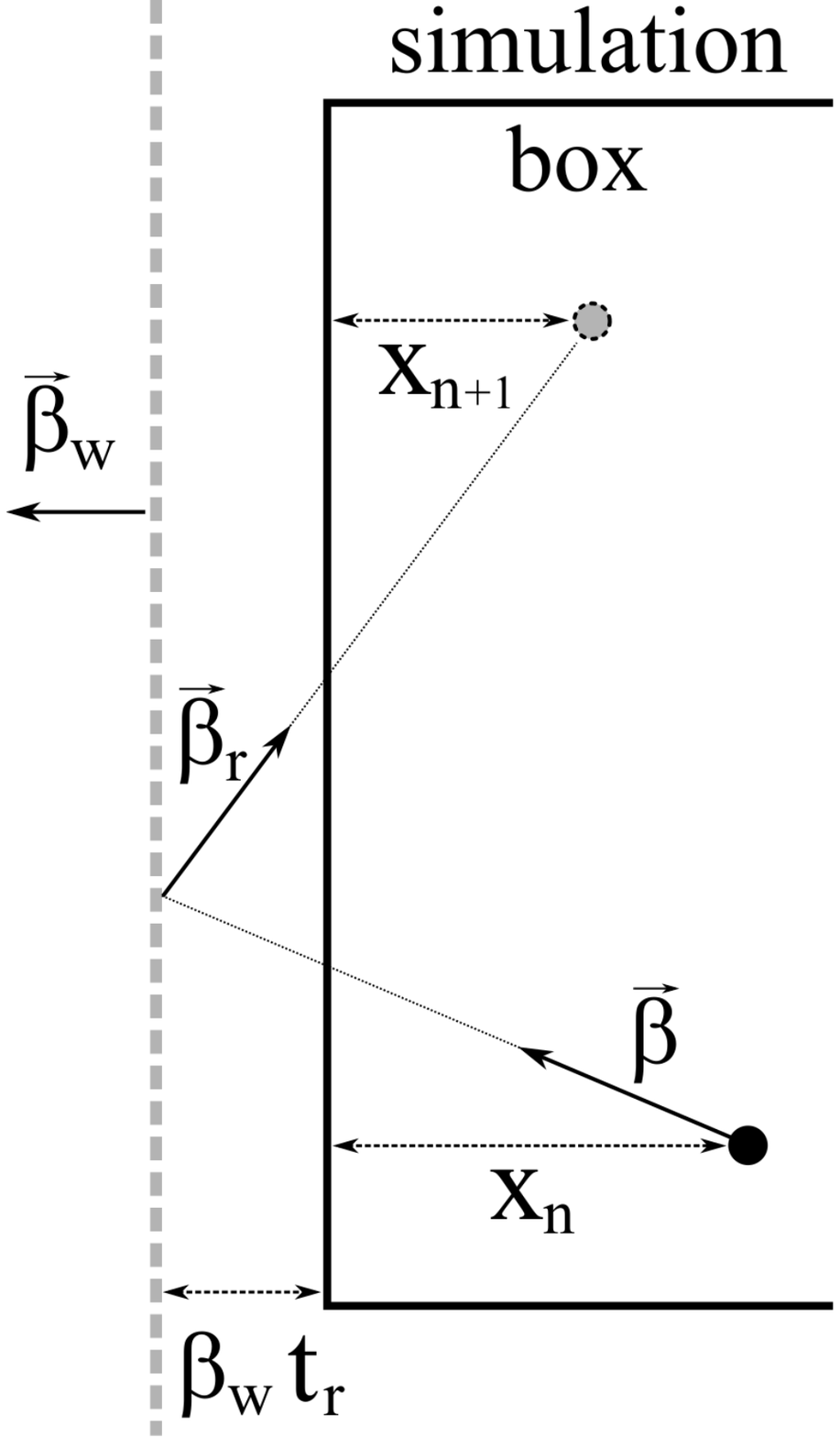}
   \caption{A schematic representation of the moving-wall boundary condition. The solid black line outlines the edges of the simulation box. At the beginning of each timestep, the moving wall (thick dashed gray line) is placed at the boundary of the simulation box and then recedes from it with constant velocity $\beta_{w}$. It reflects particles acting as a mirror in its comoving frame. Positions of a particle at the start and at the end of each timestep are marked by black and gray circles respectively, and its trajectory is shown by thin dotted line.}
   \label{fig:Moving_Wall_Boundary}
\end{figure}

The first is direct implementation. It assumes that, after leaving the simulation box through the boundary of interest, a particle reflects from the mirror that moves away from the simulation box boundary (if the particle's velocity is sufficient to catch up with the receding mirror), and then reappears inside the simulation box (if it moves to the right upon reflection and its coordinate by the end of the timestep is positive). Figure~\ref{fig:Moving_Wall_Boundary} shows the trajectory of a reflected particle. 

Consider a particle whose dimensionless energy and momentum in the simulation frame are $\gamma \equiv E/(mc^2)$ and $\vec{p} \equiv \vec{P}/(mc)$. Lorentz boost into the moving-wall frame gives $\gamma^{\prime} = \Gamma_{w} (\gamma - \beta_{w} p_{x})$ and $p^{\prime}_{x} = \Gamma_{w} (p_{x} - \beta_{w} \gamma)$. Mirror reflection from the moving wall changes sign of $p^{\prime}_{x}$, and then Lorentz boost back into the simulation frame gives
\begin{equation} \label{reflected_energy_momentum}
\begin{aligned}
    p_{x,r} & = \Gamma_{w}^2 (2 \beta_{w} \gamma - (1+\beta_{w}^2) p_{x}) \\
    \gamma_{r} & = \Gamma_{w}^2 \left( (1 + \beta_{w}^2) \gamma - 2 \beta_{w} p_{x} \right) .  \\
\end{aligned}
\end{equation}
Here $\beta_{w} \equiv v_{w}/c$ is dimensionless velocity of the reflecting wall (a signed quantity) and $\Gamma_{w} = (1-\beta_w^2)^{-1/2}$. 
Since we consider mirror reflection, the perpendicular momentum components do not change, i.e., $p_{y,r} = p_{y}$ and $p_{z,r} = p_{z}$.
The velocity of the reflected particle could be calculated as $\vec{\beta}_{r} = \vec{p}_{r} / \gamma_r$ that gives
\begin{equation} \label{reflected_velocity}
\begin{aligned}
\beta_{x,r} & = \frac{2 \beta_{w}  - \left( 1+\beta_{w}^2 \right) \beta_{x}}{ (1 + \beta_{w}^2) - 2 \beta_{w} \beta_{x}} \\
\beta_{y,r} & =  \frac{\left( 1-\beta_w^2 \right) \beta_{y}}{\left(1 + \beta_{w}^2 \right) - 2 \beta_{w} \beta_{x} } \\
\beta_{z,r} & =  \frac{\left( 1-\beta_w^2 \right) \beta_{z}}{\left(1 + \beta_{w}^2 \right) - 2 \beta_{w} \beta_{x} } \, . \\
\end{aligned}
\end{equation}
Apparently, we are interested only in particles with positive $\beta_{x,r}$; this requires positive numerator.
Any particle with positive $\beta_{x,r}$ is guaranteed to catch up with the reflecting wall; this happens at the moment
\begin{equation} \label{reflection_time}
    t_{r} = \frac{x_{n}}{\beta_{w} - \beta_{x}} \, ,
\end{equation}
where $x_{n}$ is the particle's coordinate at beginning of $n$-th timestep and $t_{r}$ is measured from the beginning of the timestep. 
The reflection occurs at coordinate $x_{r} = \beta_{w} t_{r}$.
By the end of the timestep, which lasts for $\Delta t$, the reflected particle has coordinate
\begin{equation} \label{next_position}
    \begin{split}
    x_{n+1} = x_{r} + \beta_{x,r} \left( \Delta t - t_{r} \right)
    = \\
    \frac{\left( 2 \beta_{w}  - \left( 1+\beta_{w}^2 \right) \beta_{x} \right) \Delta t  - \left( 1 - \beta_{w}^2 \right) x_{n} }
    { (1 + \beta_{w}^2) - 2 \beta_{w} \beta_{x}} \, .    
    \end{split}
\end{equation}
To return into the simulation box, the particle must satisfy $x_{n+1} \geq 0$, i.e., 
\begin{equation} \label{coordinate_selection}
    x_{n} \leq x_{n,ret} \equiv \Gamma_{w}^2 \left( 2 \beta_{w}  - \left( 1+\beta_{w}^2 \right) \beta_{x} \right) \Delta t \, .
\end{equation}
In a statistical sense, one may say that particles leaving the simulation box have $P_\mathrm{refl} = x_{n,ret} /(- \beta_{x} \Delta t)$ probability of reflection, assuming uniform distribution over $x$-coordinate and provided that $p_{x,r}>0$. This probability evaluates to
\begin{equation} \label{reflection_probability}
    P_\mathrm{refl} 
    = \frac{\Gamma_{w}^2 \left( 2 \beta_{w}  - \left( 1+\beta_{w}^2 \right) \beta_{x} \right)}{- \beta_{x}} 
    = \frac{ p_{x,r} }{ - p_{x} } \, ,
\end{equation}
where the last equality comes from comparison with Eq.~(\ref{reflected_energy_momentum}).

To summarize, the algorithm that implements the moving-wall boundary condition is the following.

1. Out of all particles leaving the simulation box, select only those whose velocity satisfies the condition
\begin{equation} \label{velocity_selection}
    \beta_{x} < \frac{2 \beta_{w} }{ 1+\beta_{w}^2 }  \, ,
\end{equation}
whereas other particles are discarded. Note that both $\beta_{w}$ and $\beta_{x}$ are negative here.

2. For every selected particle, calculate the position at the end of the timestep according to Eq.~(\ref{next_position}). 
The particles that have $x_{n+1}<0$ are discarded. 
Other particles are kept within the simulation box with new coordinates, and their energy and momentum is recalculated according to Eq.~(\ref{reflected_energy_momentum}).

Another, statistical, implementation of the moving-wall boundary condition is based on the notion that $\beta_{w}$ is intended to be the bulk velocity at the simulation box boundary, and one may consider the moving-wall reference frame as fluid-comoving frame in hydrodynamical sense. Then, the particle distribution in this frame is symmetric, $f^{\prime} \left( p^{\prime}_{x}, p^{\prime}_{y}, p^{\prime}_{z} \right) = f^{\prime} \left( - p^{\prime}_{x}, p^{\prime}_{y}, p^{\prime}_{z} \right)$. Consider particles in small neighborhood of point $\left( p^{\prime}_{x}, p^{\prime}_{y}, p^{\prime}_{z} \right)$ in the momentum space. The absolute value of $x$-component of their flux density is equal to that of the particles in the same neighborhood of the mirror-image point $\left( - p^{\prime}_{x}, p^{\prime}_{y}, p^{\prime}_{z} \right)$. Lorentz boost into the simulation-box frame transforms $p^{\prime}_{x}$ into $p_{x}$ and $-p^{\prime}_{x}$ into $p_{x,r}$ (given by Eq.~\ref{reflected_energy_momentum}). Since $x$-component of particle flux density is spatial component of a 4-vector, it transforms in the same way as $p_{x}$, which is the same component of energy-momentum 4-vector. Then, the ratio of flux densities for outgoing particles and reflected ingoing particles is given by the ratio of their momenta, $p_{x}$ and $p_{x,r}$. Thus, the reflection probability is given by Eq.~(\ref{reflection_probability}).

The algorithm for statistical implementation of the moving-wall boundary condition is the following.

1. Out of all particles leaving the simulation box, select only those whose velocity satisfies the condition from Eq.~(\ref{velocity_selection}) and discard other particles.

2. Every selected particle causes injection of the reflected particle with probability given by Eq.~(\ref{reflection_probability}). Energy and momentum of the injected particle are given by Eq.~(\ref{reflected_energy_momentum}).

{We have performed
a control test loading a boosted Maxwell-J\"{u}ttner distribution of particles (boost velocity is \(\beta_0\)), adding a consentient injection at the upstream boundary. Then run the simulation in this setup with the electromagnetic field-particle interaction disabled. With the moving wall boundary condition $\beta_{w} = \beta_0$, the flow is smooth: it remains stationary and we do not observe a persistent density pile-up, rarefaction, or pressure boundary layer. }

\section{Maxwell-J\"{u}ttner distribution}
\label{sec:MJ_distribution}
{
There are several methods for generating Maxwell-J\"{u}ttner distribution, e.g. \citet{Zenitani2015}.  Here we describe an alternative rejection-sampling method that we find convenient in practice: it uses a simple analytically invertible majorizing distribution for the momentum magnitude and then directly samples the angular distribution for a Lorentz-boosted isotropic population. It  has similarly high acceptance rate for both nonrelativistic
and relativistic temperatures.}

Let us introduce the particle's dimensionless energy (Lorentz factor) $\gamma = E/(mc^2)$ and dimensionless momentum $p=P/(mc)$, note that $p=\sqrt{\gamma^2-1}$. Let $\theta$ be the dimensionless temperature of the distribution, $\theta=kT/(mc^2)$. 

We look at generating the Maxwell-J\"{u}ttner distribution. Without the normalization factor, which is
not important for constructing the random-number generator, this distribution is given by 
\begin{equation}
f_{mj}(\gamma) = \gamma p\, \exp\left(\frac{1-\gamma}{\theta}\right),
\end{equation}
The simpler distribution 
\begin{equation}
f_{env}(\gamma) = \frac{A^2 \gamma p}{(A+p^3)^2}
\end{equation}
with appropriate choice of normalization factor $A$ can be used as majorizing function of the Maxwell-J\"{u}ttner distribution.

In the limit $\gamma\rightarrow 1$ (i.e., $p\rightarrow 0$) 
\begin{equation}
\frac{f_{env}(\gamma)}{f_{mj}(\gamma)} = 
\frac{A^2}{(A+p^3)^2} \exp\left(\frac{\gamma-1}{\theta}\right) = 
1 + \frac{p^2}{2\theta} + O(p^3)\, ,
\end{equation}
so that $f_{env}(\gamma) > f_{mj}(\gamma)$. In the limit $\gamma\rightarrow \infty$, the Maxwell-J\"{u}ttner distribution function decreases exponentially, i.e., faster than the power-law tail of $f_{env}(\gamma)$, so that $f_{env}(\gamma) > f_{mj}(\gamma)$ again. It follows from this, that one can choose the factor $A$ in such a way, that $f_{env}(\gamma) > f_{mj}(\gamma)$ everywhere in the region $(1,\infty)$ except for a single point at $\gamma_c$, where the functions have a contact of order 1. This choice of $A$ maximizes the acceptance rate.

At the point of contact both the functions $f_{env}(\gamma)$, $f_{mj}(\gamma)$ and their first derivatives are equal, that gives a set of two equations:
\begin{equation}
\left\{
\begin{array}{l}
\displaystyle
\frac{A^2}{(A+p^3)^2} = \exp\left(\frac{1-\gamma}{\theta}\right)\\[4ex]
\displaystyle
\frac{A^2}{(A+p^3)^2} \left(p + \frac{\gamma^2}{p} - \gamma^2 \frac{6 p^2}{A+p^3}\right)
= \\
\displaystyle\left(p + \frac{\gamma^2}{p} - \frac{\gamma p}{\theta}\right) 
\exp\left(\frac{1-\gamma}{\theta}\right).
\end{array}
\right.
\end{equation}
From these equations, one can separate the equation for the contact point coordinate
\begin{equation} \label{contact}
1+ 6 \gamma_c \theta -\gamma_c^2 =  6 \gamma_c \theta\, \exp\left(\frac{1-\gamma_c}{2\theta}\right)
\end{equation}
and the expression for the normalization factor
\begin{equation} \label{factor}
A = \left(1+ 6 \gamma_c \theta -\gamma_c^2\right) \sqrt{\gamma_c^2 -1}\, .
\end{equation}
In numerical applications, it makes sense to resort to use of $\epsilon = \gamma -1$ to avoid loss of precision when $\theta \ll 1$, and rewrite Eqs.~(\ref{contact}) and (\ref{factor}) accordingly. 

Equation (\ref{contact}) can be solved by the following iterative method.\\ 
1. Start with $\displaystyle \epsilon_0 = \sqrt{1+11\, \theta^2} -1 + 2\, \theta$.\\ 
2. Calculate $\displaystyle q = 3 \left(1 - \exp\left(-\frac{\epsilon_n}{2\theta}\right) \right)$.\\
3. Calculate the next approximation 
$\displaystyle \epsilon_{n+1} = \epsilon_{n} + 
\frac{\epsilon_{n}(\epsilon_{n}+2) - 2 q \theta (\epsilon_{n}+1)}{\epsilon_{n}+1 + q (2 \theta - \epsilon_{n} - 1)}$.\\
4. Go to step 2 if the desired precision is not achieved.\\
5. Calculate $A$ from Eq.~(\ref{factor}).\\
6. Multiply $A$ by a factor $(1+\delta)$ with small $\delta>0$ to allow for imprecise value of $A$. \\
Using the value of $\epsilon_3$ for $\epsilon_{c}$ in Eq.~(\ref{factor}) gives relative precision better 
than $10^{-7}$ everywhere in the range $\epsilon_c \in (0,\infty)$.

\begin{figure} 
\includegraphics[width=\columnwidth]{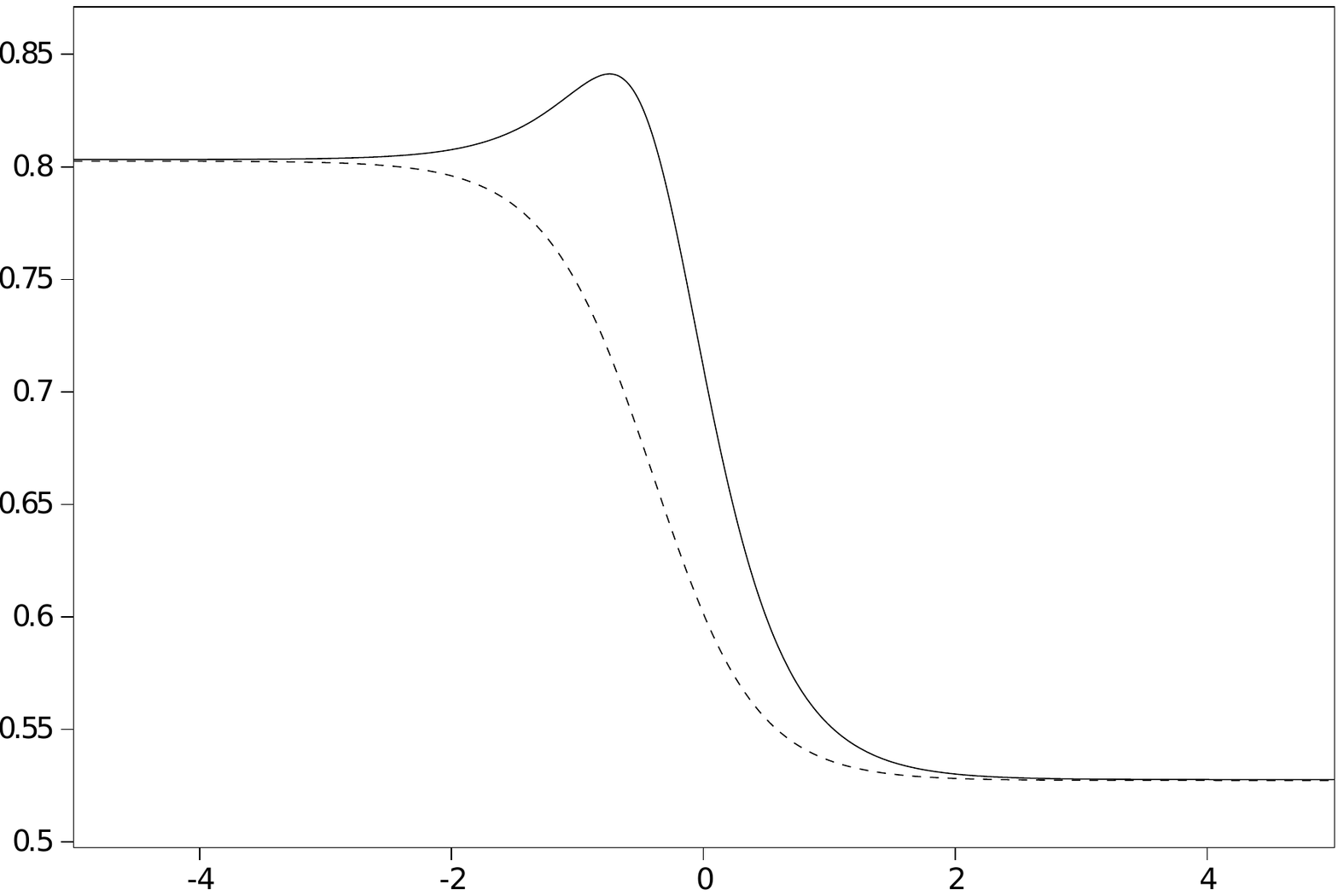}
\caption{Acceptance rate as a function of logarithm of the dimensionless temperature. Solid line is for optimal value of the normalization factor $A$, dashed line -- for approximate value of $A$ (see text for explanations).}
\label{AccRate}
\end{figure}

Alternatively, the normalization factor can be quickly estimated as 
$A = 11.38 \left(\theta^2 + 0.5534\, \theta\right)^{3/2}$ at the expense of lower acceptance rate.
The acceptance rate as a function of $\theta$, both for optimal and approximate choice of $A$, is shown in Fig.~(\ref{AccRate}).

Finally, the Maxwell-J\"{u}ttner distribution is produced by rejection sampling.\\
1. Generate two independent random numbers, $u$ and $w$, uniformly distributed in the range $[0,1)$.\\
2. Calculate $\displaystyle p = \left(\frac{A\, u}{1 - u}\right)^{1/3}$, which is distributed as 
$\displaystyle f(p) = \frac{3 A p^2}{(A+p^3)^2}$. This distribution differs from $f_{env}(\gamma)$
only by the normalization factor (note that $p\,{\rm d}p=\gamma\,{\rm d}\gamma$).\\
3. If $\displaystyle w > \exp\left( \frac{1-\sqrt{1+p^2}}{\theta}\right) \left(1+\frac{p^3}{A}\right)^2$ go to step 1.\\
4. Return $p$, which is the momentum of a random particle with the Maxwell-J\"{u}ttner distribution.

With the optimal choice of the normalization factor $A$, the acceptance rate varies from 0.803 in the non-relativistic limit ($\theta \rightarrow 0$) to 0.528 in the ultrarelativistic limit ($\theta \rightarrow \infty$), reaching maximum of 0.841 at $\theta \simeq 0.18$.

It is often desirable to generate particles in such a way that there is a macroscopic (hydrodynamical) velocity. This is equivalent to generating the particle distribution function in the (hydrodynamical) rest frame and then applying the Lorentz transformation. For isotropic distributions, it is possible to get the Lorentz boosted distribution with minor arithmetic overhead.

Given the particle's momentum norm $p$, the momentum vector components are 
\begin{equation}
\begin{array}{l}
p_{x} = p \sin (\theta) \cos (\phi),\\
p_{y} = p \sin (\theta) \sin (\phi),\\
p_{z} = p \cos (\theta).
\end{array}
\end{equation}
Particles from an isotropic 3D distribution are uniformly distributed both over $\phi$ in the range $[0,2\pi)$ and over $\mu = \cos(\theta)$ in the range $[-1,1]$.

Without loss of generality we can assume that the hydrodynamic velocity is collinear with the $z$-axis.
In the reference frame, where the hydrodynamic velocity is $\beta_0$ (and the hydrodynamic Lorentz factor is  $\gamma_0$), the transverse components of the momentum vector ($p_{x}$ and $p_{y}$) do not change, so that the Lorentz boosted distribution remains uniform over $\phi$. To find the distribution over $\mu$, we note that the number density is the timelike component of the particle-number four-current, and therefore it transforms as the timelike component of the particles' four-velocity, i.e., 
\begin{equation}
n^{\prime} \propto \gamma^{\prime} = \gamma_0 \gamma + \beta_0 \gamma_0 p_{z} = \gamma_0 \gamma (1+ \beta_0 \beta \mu),
\end{equation}
where $\beta = p/\gamma$ is the particle velocity.

Thus, the problem of generating a Lorentz boosted distribution is reduced to the problem of generating the distribution 
\begin{equation}
f^{\prime}(\mu) \propto 1+ \beta_0 \beta \mu 
\end{equation}
instead of uniform distribution over $\mu$ in the hydrodynamical rest frame. 

The full algorithm for generation of a Lorentz boosted distribution is the following.\\
1. Generate the particle's total momentum $p$ from any isotropic (in the hydrodynamical rest frame) distribution.\\
2. Generate a random number $u$ uniformly distributed in the range $[0,1]$.\\
3. Calculate $\displaystyle \mu=\frac{\sqrt{(1 - \beta_0 \beta)^2 + 4 u \beta_0 \beta} - 1}{\beta_0 \beta}$, which is distributed as $\displaystyle f(\mu)=\frac{1 + \beta_0 \beta \mu}{2}$\\
4. Calculate $p_{z}=\mu p$.\\
5. Generate another independent random number $w$ uniformly distributed in the range $[0,1)$.\\
6. Calculate $p_{x}= \sqrt{1-\mu^2}\, \cos (2\pi w)\, p$ and $p_{y}= \sqrt{1-\mu^2}\, \sin (2\pi w)\, p$.\\

\bibliographystyle{apsrev4-2}
\bibliography{shocks}

\end{document}